\documentclass[aps,prd,twocolumn,groupedaddress]{revtex4-1}

 \pdfoutput=1

\usepackage{tensor}
\usepackage[pdftex]{graphicx}
\usepackage{amscd}
\usepackage{amsthm}
\usepackage{amssymb}
\usepackage{combelow}

\usepackage{bm}
\usepackage{amstext}
\usepackage{psfrag}
\usepackage{amsmath}
\usepackage{amsfonts}
\usepackage{units}
\usepackage{mathrsfs}
\newcommand{\p}{\partial}

\newcommand{\dd}{{\rm d}}

\newcommand{\bd}{\begin{definition}}                
\newcommand{\ed}{\end{definition}}                  
\newcommand{\bc}{\begin{corollary}}                 
\newcommand{\ec}{\end{corollary}}                   
\newcommand{\bl}{\begin{lemma}}                     
\newcommand{\el}{\end{lemma}}                       
\newcommand{\bp}{\begin{proposition}}            
\newcommand{\ep}{\end{proposition}}                
\newcommand{\bere}{\begin{remark}}                  
\newcommand{\ere}{\end{remark}}                     

\newcommand{\bt}{\begin{theorem}}
\newcommand{\et}{\end{theorem}}

\newcommand{\be}{\begin{equation}}
\newcommand{\ee}{\end{equation}}

\newcommand{\bit}{\begin{itemize}}
\newcommand{\eit}{\end{itemize}}
\newtheorem{theorem}{Theorem}[section]
\newtheorem{corollary}[theorem]{Corollary}
\newtheorem{lemma}[theorem]{Lemma}
\newtheorem{proposition}[theorem]{Proposition}
\theoremstyle{definition}
\newtheorem{definition}[theorem]{Definition}
\theoremstyle{remark}
\newtheorem{remark}[theorem]{Remark}

\begin{document}


\title{Affine sphere spacetimes which satisfy the relativity principle}


\author{E. Minguzzi}
\email[]{ettore.minguzzi@unifi.it}
\affiliation{Dipartimento di Matematica e Informatica ``U. Dini'', Universit\`a
degli Studi di Firenze, Via S. Marta 3,  I-50139 Firenze, Italy.}


\date{\today}

\begin{abstract}In the context of Lorentz-Finsler spacetime theories  the relativity principle holds at a spacetime point if the indicatrix (observer space) is homogeneous.
 We point out that in four spacetime dimensions there are just three kinematical models which respect an exact form of the relativity principle and for which all observers agree on the spacetime volume. They have necessarily affine sphere indicatrices. For them every observer which looks at  a flash of light emitted by a point would observe, respectively, an expanding (a) sphere, (b) tetrahedron, or (c)  cone, with barycenter at the point. The first model corresponds to Lorentzian relativity, the second one has been studied by several authors though the relationship with affine spheres passed unnoticed, and  the last one has not been previously recognized and it is studied here in some detail. The symmetry groups are $O^+(3,1), \mathbb{R}^3, O^+(2,1)\times \mathbb{R}$, respectively. In the second part, devoted to the general relativistic theory, we show that the field equations can be obtained by gauging the Finsler Lagrangian symmetry while avoiding  direct use of Finslerian curvatures.
 We construct some notable affine sphere spacetimes which in the appropriate velocity limit return the Schwarzschild, Kerr-Schild, Kerr-de Sitter,  Kerr-Newman, Taub,  and FLRW spacetimes, respectively.
\end{abstract}


\maketitle

{\small
\setcounter{tocdepth}{3}
\tableofcontents
}
\section{Introduction}

Finslerian modifications of general relativity have received renewed attention in recent years. Theoretically  they share with general relativity the whole edifice of causality theory including the  celebrated singularity theorems \cite{minguzzi13d,minguzzi15}, a result which does not seem to be shared by any other alternative gravity theory.

Observations are also suggesting that we consider  these theories, for they seem to  provide the correct mathematical framework for the study  of the low-$\ell$ anisotropy of the CMB temperature \cite{pavlov10,chang13b}.

Finslerian proposals have been advanced in order to explain some anisotropic features of the universe, including the observed anisotropy in the galaxy bulk flow \cite{chang13}, and they can  also have a role in the dynamics of  dark energy and dark matter \cite{basilakos13,chang13c}.

Finslerian modifications of gravity and of particle dynamics are in fact quite ubiquitous even at the quantum level, due to the fact that modified dispersion relations often lead to geometries of Finslerian type \cite{girelli07,amelino-camelia14,itin14}.


This work is devoted to the study of four-dimensional Finslerian spacetimes which satisfy the relativity principle. The adjective `Finslerian' means that no assumption on the isotropy of the speed of light will be made.
We shall obtain Finslerian generalizations of the notable spacetimes of Einstein's gravity including Schwarzschild's.


Finslerian generalizations of say, the Schwarzschild or of the Friedman metric
have long been sought. Most proposals  \cite{asanov92,rutz93,stavrinos08,silagadze11,basilakos13,lammerzahl12,li14,rahaman15},  have used  one of the following ingredients (a) Randers metrics, (b) direct sum metrics, (c) perturbation.
  Instead, we shall impose the relativity principle at every point showing that this condition restricts significatively the geometry of the indicatrix. For a particular conic anisotropic geometry we will be able to obtain, almost unambiguously, the Finslerian generalization of the notable general relativistic metrics from  the mentioned requirement of relativistic invariance and the imposition of a suitable general relativistic limit for low velocities.

  Although we do not impose dynamical equations, it is likely that  these spacetimes could be obtained as exact solutions of the sought for gravitational Finsler equations. In fact they could possibly be used to  identify them. Historically, it has often been the case that exact solutions, by respecting symmetry and other requirements, have been found before the field equations (e.g.\ the Coulomb field was determined long before Maxwell's equations).

  Let us introduce some notations in order to be more specific.


In Finslerian generalizations of general relativity the spacetime is a $n+1$-dimensional manifold endowed with a Finsler Lagrangian $\mathscr{L}\colon \Omega \to \mathbb{R}$, $\Omega\subset TM\backslash \{0\}$, where $\Omega$ is an open sharp  convex cone subbundle of the slit tangent bundle, $\mathscr{L}$ is positive homogeneous of degree two, that is $\forall s>0, y\in \Omega_x$, $\mathscr{L}(x,sy)=s^2 \mathscr{L}(x,y)$,  $\mathscr{L}$ is negative on $\Omega$ and converges to zero at the boundary $\p \Omega$, and finally, the fiber Hessian $g_{\mu \nu}=\p^2\mathscr{L}/\p y^\mu \p y^\nu$ is Lorentzian. We shall not demand $\mathscr{L}$ to be differentiable at the boundary $\p \Omega$, namely we adopt the {\em rough model} discussed in \cite{minguzzi14h}. The set $\Omega_x$ represents the set of future directed timelike vectors at $x\in M$.

The indicatrix  $\mathscr{I}_x\subset \Omega_x$ is the  locus where $2\mathscr{L}=-1$ and it represents the  velocity space of observers (this is the usual hyperboloid in general relativity).
By positive homogeneity the Finsler Lagrangian can be recovered from the indicatrix as follows,
for $y\in \Omega_x$
\begin{equation} \label{diw}
\mathscr{L}(x,y)=-s^2/2, \ \textrm{ where }  s \textrm{ is such that } \ y/s\in \mathscr{I}_x.
\end{equation}
By positive homogeneity the formulas $\mathscr{L}=\frac{1}{2} g_{\mu \nu} y^\mu y^\nu$, $\frac{\p \mathscr{L}}{\p y^\mu}=g_{\mu \nu} y^\nu$, hold true, where the metric might depend on $y$. If it is independent of $y$ then we are in the quadratic case which corresponds to Lorentzian geometry and general relativity.
The Cartan torsion is $C_{\mu \nu \alpha}=\frac{1}{2} \frac{\p}{\p y^\alpha} g_{\mu \nu}$. It is symmetric and annihilated by $y^\mu$. The mean Cartan torsion is its contraction
\begin{equation} \label{aqs}
I_\alpha:=g^{\mu \nu} C_{\mu \nu \alpha}=\frac{1}{2}\frac{\p}{\p y^\alpha} \log \vert \det g_{\mu \nu}\vert .
\end{equation}

In a series of recent works we have stressed the importance of the  Lorentz-Finsler spaces for which $I_\alpha=0$, which we termed {\em affine sphere spacetimes} \cite{minguzzi14c,minguzzi15d,minguzzi15e}. Indeed, these spaces have hyperbolic affine sphere indicatrices and a well defined volume form  independent of the fiber coordinates. Their importance stems from the fact that  affine sphere spacetimes are in one-to-one correspondence with   pairs given by (a) a distribution of sharp cones over $M$ and (b) a volume form on $M$. This property shows that affine sphere spacetimes reflect the notions of measure and order on spacetime \cite{minguzzi15e}.

In what follows we recall the construction and the interpretation of the general theory as developed in \cite{minguzzi15e}.
Let $\{x^\alpha\}$ be local coordinates on $M$ and let  $\{x^\alpha,y^\alpha\}$ be the induced local coordinates on $TM$. We shall be mostly interested on a single tangent space $T_xM$ so we shall often omit the dependence on $x$.

The indicatrix at $y\in \mathscr{I}_x$ is everywhere transversal to $y$. It is particularly convenient to regard the indicatrix as the image of an embedding
\[
f\colon {\bf v}\to y=-\frac{1}{u({\bf v})} (1,{\bf v})
\]
where ${\bf v}=y^i/y^0$, for a function $u({\bf v})$ called {\em Lagrangian} (actually it is the Lagrangian per unit mass). The relationship with the  Finsler or Super Lagrangian $\mathscr{L}$ is given by
\begin{align}
\mathscr{L}((y^0,{\bf y}))&=-\frac{1}{2} (y^0)^2 u^2({\bf y}/y^0), \label{osd}\\
u({\bf v})&=-\sqrt{-2\mathscr{L}((1,{\bf v}))}. \label{sop}
\end{align}
The Hamiltonian (per unit mass) is given by the Legendre transform of $u$,  $u^*({\bf p})$. The embedding ${\bf p} \mapsto (-u^*({\bf p}),{\bf p})$ is an affine sphere in $T^*_xM$ asymptotic to the polar cone $\Omega_x^*$. Sometimes it is convenient to consider the Legendre transform $\mathscr{H}$ of $\mathscr{L}$. It is called Finsler Hamiltonian and some of its properties are investigated in \cite{minguzzi13c}.

We say that $\{y^\alpha\}$ are {\em observer coordinates} if the Taylor series expansion of $u$ has the classical form $u=-1+\frac{{\bf v}^2}{2}+o(\vert {\bf v}\vert^2)$. It can be shown \cite{minguzzi15e} that for every point on the indicatrix $\hat y\in \mathscr{I}$ there are observer coordinates such that ${\hat y}=(1,0,0,0)$. Observer coordinates can also be characterized by this condition and by $g_{\mu \nu}(\hat y)=\eta_{\mu \nu}$ where $\eta$ is the Minkowski metric.

The vector ${\bf v}$ represents then the velocity of a test particle as seen from the observer and it belongs to a convex set $D_{\hat y}:=\{{\bf v}\colon u<0\}$, which represents the velocity domain of massive particles as seen from the observer ${\hat y}$. The domain for the phase velocity ${\bf p}/u^*({\bf p})$ is given by the dual of $D_{\hat y}$, $D^*_{\hat y}$, and observer coordinates can be characterized equivalently by the condition that the expansion of $u^*({\bf p})$ is $u^*=1+\frac{{\bf p}^2}{2}+o(\vert {\bf p}\vert^2)$, namely that the dispersion relation for massive particles should reduce to the classical one in the appropriate limit of low velocity.

The previous definitions and concepts make sense in any Lorentz-Finsler spacetime.
We have an {\em affine sphere spacetime} if at every event the indicatrix is a hyperbolic affine sphere, or equivalently, if the mean Cartan torsion vanishes, $I_\alpha=0$.
The indicatrix is an affine sphere if and only if $u$ satisfies a Monge-Amp\`ere equation which in observer coordinates of observer $\hat y$ takes the very simple form
\begin{equation} \label{mon}
\det u_{ij}=\Big(-\frac{1}{u}\Big)^{n+2}, \qquad u\vert_{\p D_{\hat y}}=0
 \end{equation}
 Actually, this equation holds  in arbitrary coordinates $\{y'^\alpha\}$ provided the coordinate change between observer coordinates $y^\alpha$ and $y'^\alpha$ is linear and unimodular (unit determinant).

Our next step is to introduce the concept of {\em relativity principle}. We mentioned that $\Omega_x$ represents the sets of timelike vectors and that we need a hypersurface (indicatrix) $\mathscr{I}_x$ inside $\Omega_x$ and asymptotic to the boundary $\p \Omega_x$ in order to define the observer space (and hence the Finsler Lagrangian through (\ref{diw})). On $T_xM$ acts the group of unimodular linear transformations.
 We say that the relativity principle holds true if there is a transitive action on $\mathscr{I}_x$ by a  subgroup  $G$ of the unimodular linear group. This transitive action expresses the fact that all observers are kinematically equivalent, namely that they cannot determine their position on the velocity space by means of local measurements probing its geometry. The unimodularity condition is there to guarantee that all observers will agree on the spacetime volume form. Of course, for the usual general relativistic spacetimes the indicatrix is the hyperboloid $\mathbb{H}$, the timelike cone is round and $G$ is nothing but the Lorentz group, cf.\ Sec.\ \ref{sta}.

If we add the dilatations to  $G$ we get a group $\mathbb{R}^+\times G$ which by acting transitively on $\Omega_x$ shows that $\Omega_x$ itself is a homogeneous cone. Now, every sharp convex cone admits, up to dilatations, a unique affine sphere asymptotic to it (Cheng-Yau theorem), which for the case of homogeneous cones coincides with a level set of the {\em characteristic function of  the cone} \cite{vinberg63,sasaki80}. This hypersurface is the  only hypersurface which is invariant under the action of $G$ where $\mathbb{R}^+\times G$ is the automorphism group of the cone, and $G$ is the unimodular factor.

In other words every spacetime which satisfies the relativity principle according to our definition has homogeneous (timelike) cones and indicatrices which are affine spheres. Thus they are particular instances of affine sphere spacetimes. Equivalently, a spacetime satisfies the relativity principle if and only if it is an affine sphere spacetime and the domains $D_{\hat y}$ do not depend on $\hat y$ (up to space rotations). Namely, all observers agree on the dependence of the speed of light on direction.

Fortunately, homogeneous cones have been classified \cite{rothaus63,vinberg63,faraut94}, a fact which implies a classification of homogeneous hyperbolic affine spheres. For any dimension there are just a few homogeneous cones.
Therefore, it is  of interest to study  those four dimensional affine sphere spacetimes which satisfy the relativity principle.




\begin{remark}
We stress that the  homogeneity of the cone does not guarantee that the  relativity principle is satisfied since the indicatrix must also be an affine sphere. For instance, the Finsler Lagrangian of Example 1 in \cite{minguzzi13c} has the same round light cone of Minkowski spacetime but does not satisfy the relativity principle since its indicatrix is not an affine sphere (i.e.\ the function $u$ associated to the Finsler Lagrangian does not satisfy  Eq.\ (\ref{mon}) above). In fact, we know that Eq.\ (\ref{mon}) above has a unique solution, which  for round cones is that of Minkowski spacetime.

We mention that the relativity principle could be generalized dropping the unimodularity condition for the transitive group. In this case the indicatrix would not be  an affine sphere.

While the relativity principle restricts very much the geometry of the cone, there are plenty of affine sphere spacetimes which do not satisfy it. It is sufficient to take any distribution of convex cones obtained perturbing slightly the isotropic cones of a general relativistic spacetime so as to get a distribution of non-round cones. The affine sphere indicatrices inside the cones and then the Finsler Lagrangian are uniquely  determined by Eq.\ (\ref{mon}).

\end{remark}

\section{ The special theory}

In this section we  restrict ourselves to the preliminary case in which $\mathscr{L}$ does not depend on $x$.

\subsection{Theories which satisfy the relativity principle} \label{exe}

In Lorentz-Finsler geometry the indicatrix is asymptotic to the  cone of lightlike vectors. The metric induced on the indicatrix has to be definite, due to the  Lorentzianity of the vertical Finsler metric,  and since it coincides with the equiaffine metric (see e.g.\ \cite{laugwitz57,minguzzi14h}), the indicatrix is a definite  hypersurface in the sense of affine differential geometry (namely locally strongly convex).  We are interested in those three dimensional hypersurfaces $N$ which are locally homogeneous, namely  for every $p, q\in N$ there are neighborhoods $U_p, U_q$ and a unimodular bijective affine map from  $U_p$ to $U_q$. Since these hypersurfacces have to be asymptotic to a sharp cone, by the classification given in \cite{dillen93}, they are necessarily hyperbolic affine spheres.

Mathematicians have long investigated the classification of homogeneous cones and consequently that of homogeneous affine spheres \cite{vinberg63}. In  a four-dimensional affine space \cite{dillen93}  there are only three possible  locally homogeneous hyperbolic affine spheres which we interpret and study in Sections \ref{sta}, \ref{stb} and \ref{stc}, giving the expressions of the Lagrangian in observer coordinates. Their associated cones are actually self-dual, namely linearly isomorphic with the dual cone. It must be recalled here that a cone is {\em reducible} if it is the Cartesian product of lower dimensional cones.
In dimension 4 or less the only irreducible homogeneous cones are necessarily self-dual and are given by the half-line of positive real numbers $\mathbb{R}^+$, which is of course one-dimensional, and by the Lorentz cones of dimension 3 and 4 (the Lorentz cone of dimension 2 is reducible). Other reducible  (self-dual) homogeneous cones can be obtained by multiplying irreducible (self-dual) homogeneous cones. As a consequence, the above three  mentioned cases are really obtained from the product of round cones, an operation which at the level of the indicatrices is called Calabi product \cite{calabi72}.

We have observed that in four spacetime dimensions there are only three possible
hyperbolic affine sphere indicatrices which are homogeneous. Let us study and interpret them finding their expression in observer coordinates.


\subsubsection{Isotropic relativity}\label{sta}
Let us consider the usual velocity space of special and general relativity, namely the hyperboloid $H^n$: $y^0=\sqrt{1+{\bf y}^2}$. In the Lorentzian spacetime of general relativity it  is obtained by selecting at $T_xM$ an orthonormal basis for which $e_0$ is timelike.
The parametrization $y=-\frac{1}{u({\bf v})}(1,{\bf v})$ holds with
\[
u=-\sqrt{1-{\bf v}^2} ,
\]
where the domain of the velocity is determined by the condition $u<0$ thus it is a sphere centered at the origin
\[
D=\{{\bf v}\colon \Vert {\bf v}\Vert< 1\}.
\]
 As the domain is a sphere, the speed of light is isotropic.
We have
\[
u_i=\frac{v^i}{\sqrt{1-{\bf v}^2}}, \qquad u_{ij}=\frac{1}{\sqrt{1-{\bf v}^2}} \Big(\delta_{ij}+\frac{v^i v^j}{1-{\bf v}^2}\Big),
\]
which shows that $u_{ij}$ is positive definite. By the rank one update determinant formula $\det u_{ij}=(1-{\bf v}^2)^{-\frac{n+2}{2}}=(-\frac{1}{u})^{n+2}$. We have just checked that the indicatrix is an affine sphere.
The Finsler Lagrangian is (Eq.\ (\ref{osd}))
\[
\mathscr{L}=\frac{1}{2}\big(-(y^0)^2+{\bf y}^2\big),
\]
and the Finsler metric is the usual Minkowski metric $g_{\alpha \beta}(y)=\eta_{\alpha \beta}$, where $\eta_{\alpha j}=\delta_{\alpha j}$ and $\eta_{00}=-1$. The timelike cone is $\Omega=\{y\in T_xM: y^0>\Vert{\bf y}\Vert\}$. The affine sphere $H^n$ is homogeneous and the transitive symmetry group is the isochronous Lorentz group $O^+(3,1)$.

Concerning the dual formulation, since $u=-\sqrt{1-{\bf v}^2}$ we have ${\bf p}=\frac{{\bf v}}{\sqrt{1-{\bf v}^2}}$ and
\[
u^*({\bf p})=\sqrt{1+{\bf p}^2}\left(=\!\frac{1}{\sqrt{1-{\bf v}^2}}\right), \quad \mathscr{H}=\frac{1}{2}\big(-p_0^2+{\bf p}^2\big).
\]
Observe that the phase velocity coincides with the (group) velocity.

\subsubsection{The tetrahedral  anisotropic theory} \label{stb}
In this section we study a tetrahedral anisotropic model which satisfies the relativity principle.
G.\ \cb{T}i\cb{t}eica for $n=2$ and E.\ Calabi \cite{calabi72} for general $n$ have shown that the set
\begin{equation}
\mathscr{I}_x=\{ y\colon \ \tilde y^0 \tilde y^1\cdots \tilde y^n=(n+1)^{-\frac{n+1}{2}}, \ \tilde y^\alpha>0\},
\end{equation}
is a hyperbolic homogeneous affine sphere. It is the Calabi product of zero-dimensional hyperbolic affine spheres.
 Its timelike cone is the positive quadrant $\Omega_x=\{y\colon \tilde y^\alpha >0\}$ thus the light cone is not $C^1$ and is not strictly convex. Its section is affinely equivalent to a simplex $\Delta^n$. Observe that the $\tilde y^0$-axis is lightlike  (it belongs to the boundary of $\p \Omega_x$) thus the point $(1,0,0,0)$ does not belong to the indicatrix and hence the coordinates are not observer coordinates.
  Still all the formalism can be used to check whether it is really an affine sphere.
 The coordinates of an observer are  linearly related with $\{\tilde y^\alpha\}$ and will be given in a moment. In Calabi coordinates the domain  $\tilde D=\{{\bf \tilde v}: \tilde v^i>0\}$ is non-compact and
\begin{equation} \label{led}
\tilde u=-(n+1)^{1/2}(\tilde v^1 \tilde v^2\cdots \tilde v^n)^{1/(n+1)} .
\end{equation}
The partial derivatives are
\begin{align*}
\tilde u_i=\frac{\tilde u}{(n+1) \tilde v^i}, \ \ \tilde u_{ij}=-\frac{\tilde u}{(n\!+\!1) (\tilde v^i)^2} \, \delta_{ij}+\frac{\tilde u}{(n\!+\!1)^2\, \tilde v^i \tilde v^j} ,
\end{align*}
thus $\det \tilde u_{ij}=(-\frac{1}{ \tilde u})^{n+2}$ and by Eq.\ (\ref{mon})  $\mathscr{I}_x$ is a hyperbolic affine sphere.
The Finsler Lagrangian is
\begin{equation} \label{cal}
\mathscr{L}_C=-\frac{n+1}{2} \, (\tilde y^0 \tilde y^1\tilde y^2 \cdots \tilde y^n)^{\frac{2}{n+1}}.
\end{equation}
This Lagrangian was also considered by  Berwald and Mo\'or \cite{berwald39,moor54} and it has been investigated in several mathematical and physical works, e.g.\ \cite{matsumoto78b,asanov85,neagu11,balan10,balan12}.


Bogoslovsky and Goenner \cite{bogoslovsky98,bogoslovsky99}    considered the next Lagrangian (for the physical case $n=3$) to which they arrived through symmetry considerations unrelated to the theory of affine spheres
\begin{align*}
\mathscr{L}_{BG}&=-\frac{1}{2}\big[ (y^0-y^1-y^2-y^3)^{(1+a+b+c)/2}  \\& \qquad \quad (y^0-y^1+y^2+y^3)^{(1+a-b-c)/2}\\& \qquad \quad (y^0+y^1-y^2+y^3)^{(1-a+b-c)/2}\\& \qquad \quad(y^0+y^1+y^2-y^3)^{(1-a-b+c)/2} \big] ,
\end{align*}
where all the exponents are demanded to be positive.
We have calculated the determinant of the spacetime metric
\begin{align*}
\det g_{\alpha \beta}=&-\big(a^4-2 a^2 \left(b^2+c^2+1\right)+8 a b c+b^4  \\
&  \qquad -2 b^2 \left(c^2+1\right)+\left(c^2-1\right)^2\big) \\& \quad (y^0-y^1-y^2-y^3)^{2 (a+b+c)} \\& \quad (y^0+y^1-y^2+y^3)^{-2 (a-b+c)} \\& \quad (y^0-y^1+y^2+y^3)^{2 (a-b-c)}  \\& \quad (y^0+y^1+y^2-y^3)^{-2 (a+b-c)}.
\end{align*}
The first parenthesis has to be non-zero for the metric to be non-degenerate. As a consequence the determinant depends on $y$ unless all the exponents vanish which implies $a=b=c=0$. For this choice the Lagrangian is just Calabi's up to a linear change of coordinates (such that $\det \p \tilde y/\p y=1$), thus the indicatrix is a known hyperbolic affine sphere. In this case we have $\det g_{\alpha \beta}=-1$.

Let us consider the Calabi Lagrangian in the coordinates by Bogoslovsky and Goenner
\begin{align}
\begin{split} \label{mod}
\mathscr{L}_C&=-\frac{1}{2}\big[ (y^0-y^1-y^2-y^3)^{1/2} (y^0-y^1+y^2+y^3)^{1/2}  \\& \qquad \quad (y^0+y^1-y^2+y^3)^{1/2}(y^0+y^1+y^2-y^3)^{1/2} \big] ,
\end{split}
\end{align}
The vector $\hat y=(1,0,0,0)$ belongs to the indicatrix and a calculation shows that at this point
$g_{\alpha \beta}=\eta_{\alpha \beta}$, thus $\{ y^\alpha\}$ coincides with the coordinate system chosen by the observer ${\hat y}$ according to the general theory previously illustrated.
The Cartan torsion at the same point has, up to symmetries, the only non-vanishing component  $C_{123}=1$. The Cartan curvature has, up to symmetries and at the same point, the only non-vanishing components $C_{0123}=-1$, $C_{iijj}=2$ for $i,j=1,2,3$. The function $u$ is
\begin{align*}
u&=\!-\big[(1\!-\!v_1\!-\!v_2\!-\!v_3)(1\!-\!v_1\!+\!v_2\!+\!v_3) \\ & \qquad \qquad \qquad \qquad (1\!+\!v_1\!-\!v_2\!+\!v_3)(1\!+\!v_1\!+\!v_2\!-\!v_3)\big]^{1/4} .
\end{align*}


Bogoslovsky and Goenner have also shown that their Lagrangian is invariant under a certain group of symmetries \cite{bogoslovsky99} which, however, do not have unit determinant.
 As a consequence, in Bogoslovsky and Goenner's theory observers  cannot agree on the spacetime volume.
 For $a=b=c=0$ there is no such difficulty since the indicatrix is the
  Calabi affine sphere, which is well known to be  homogeneous  \cite{calabi72}. Calabi has shown that the symmetry group is the commutative group $\mathbb{R}^n$, thus it has the minimal dimension for a transitive action on an $n$-dimensional manifold. Its action is for $\alpha_i\in \mathbb{R}$
\begin{equation} \label{nok}
\tilde{y}^i \mapsto e^{\alpha_i} \tilde{y}^i \quad (\textrm{no sum over } i),  \quad \tilde y^0\mapsto e^{-\sum_i\alpha_i} \tilde y^0 .
\end{equation}

\begin{figure}[ht!]
\centering
 \includegraphics[width=5cm]{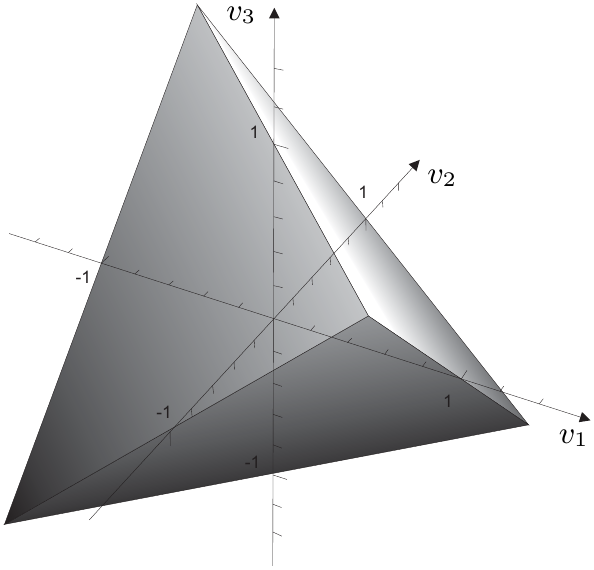}
 \caption{The velocity space for the tetrahedral anisotropic model (Sec.\ \ref{stb}).} \label{max}
\end{figure}

If some of the constants $a,b,c$ do not vanish the Bogoslovsky and Goenner's indicatrix is homogeneous but it is not an affine sphere. These authors have given a nice picture of the velocity domain $D$  \cite{bogoslovsky98},
\begin{align*}
D&=\{{\bf v}: v_1+v_2+v_3<1, v_1-v_2-v_3<1,\\
& \qquad \quad \ v_2-v_1-v_3<1, v_3-v_1-v_2<1\}.
\end{align*}
 It is a tetrahedron centered at the origin and is independent of the constants $a,b,c$ (see Fig.\ \ref{max}).

Let us come to the dual formulation.
Let us consider the Calabi Lagrangian in arbitrary dimension, Eq.\ (\ref{cal}).
The Finsler Hamiltonian is
\begin{equation}
\mathscr{H}_C=-\frac{n+1}{2} \, (\tilde p_0 \tilde p_1\tilde p_2 \cdots \tilde p_n)^{\frac{2}{n+1}}.
\end{equation}
Since $\tilde u$ is given by (\ref{led})
we have $\tilde p_i=\tilde u_i=\frac{\tilde u}{(n+1) \tilde v^i}$, and
the Legendre transform is
\[
\tilde u^*=-\frac{1}{n+1}\, \tilde u=\Big[\frac{-1}{(n+1)^{1/2}}\Big]^{n+1} \frac{1}{\tilde p_1 \tilde p_2 \cdots \tilde p_n}.
\]
The interpretation of this formula is not straightforward since these are not observer coordinates (hence the tilde).

Let us consider the case $n=3$ in observer coordinates, namely the Finsler Lagrangian (\ref{mod}). We have from Eq.\ (\ref{sop})
\begin{align*}
u&= (1-v_1-v_2-v_3)^{1/4} (1-v_1+v_2+v_3)^{1/4}  \\
& \quad \ (1+v_1-v_2+v_3)^{1/4}(1+v_1+v_2-v_3)^{1/4} .
\end{align*}
The Legendre transform is
\[
u^*=u^{-3/4}({\bf v}^2+2 v_1v_2v_3-1).
\]
We have not been able to write it in terms of ${\bf p}$. The Finsler Hamiltonian is
\begin{align}
\begin{split}
\mathscr{H}_C&=-\frac{1}{2}\big[ (-\!p_0\!-\!p_1\!-\!p_2\!-\!p_3)^{1/2} (-\!p_0\!-\!p_1\!+\!p_2\!+\!p_3)^{1/2}  \\& \qquad \quad (-\!p_0\!+\!p_1\!-\!p_2\!+\!p_3)^{1/2}(-\!p_0\!+\!p_1\!+\!p_2\!-\!p_3)^{1/2} \big] . \nonumber
\end{split}
\end{align}



\subsubsection{The conical anisotropic theory}\label{stc}
\begin{figure}[ht!]
\centering
 \includegraphics[width=6cm]{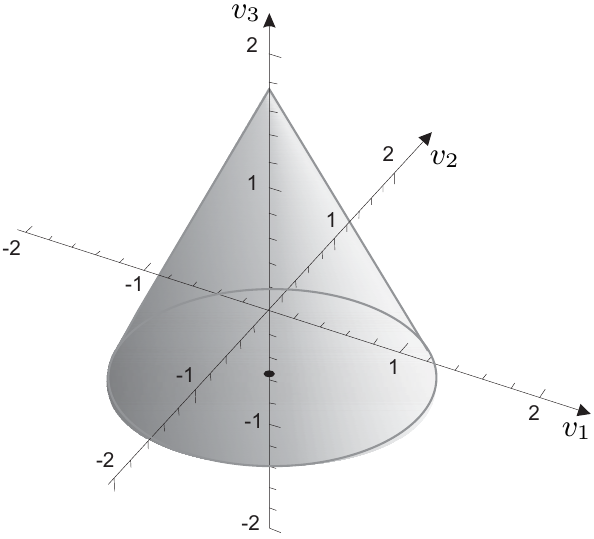}
 \caption{The velocity space for the conical anisotropic model  (Sec.\ \ref{stc}).} \label{min}
\end{figure}

In this section we study a  conical anisotropic model which respects the relativity principle.
We are going to consider a homogeneous hyperbolic affine sphere indicatrix which is a Calabi product between  0-dimensional and  2-dimensional hyperbolic affine spheres.
In suitable coordinates
the Finsler Lagrangian is
\begin{equation} \label{lod}
\mathscr{L}=-\frac{2}{3^{3/4}} (\tilde y^3)^{1/2} [(\tilde y^0)^2-(\tilde y^1)^2-(\tilde y^2)^2]^{3/4}.
\end{equation}
The indicatrix is
\begin{equation}
(\tilde y^3)^2[(\tilde y^0)^2-(\tilde y^1)^2-(\tilde y^2)^2]^3=3^3/4^4.
\end{equation}
 Let us write this Lagrangian in observer coordinates as presented in the Introduction.
The coordinate change is a rotation of $30^\circ$ (thus $\det \p \tilde y/\p y=1$ and $\rho=1$ as expected)
\begin{equation} \label{bot}
\tilde y^0=\frac{\sqrt{3}}{2}\, y^0-\frac{1}{2}\, y^3, \ \tilde y^1=y^1, \ \tilde y^2=y^2, \ \tilde y^3= \frac{1}{2}\,y^0+\frac{\sqrt{3}}{2}\, y^3,
\end{equation}
thus the Lagrangian is
\begin{align}
\begin{split}
\mathscr{L}&=-\frac{2}{3^{3/4}} \left(\frac{1}{2}\, y^0+\frac{\sqrt{3}}{2} \,  y^3\right)^{\!1/2} \\ &\qquad \ \left(\left(\frac{\sqrt{3}}{2}\, y^0-\frac{1}{2}\, y^3\right)^2-(y^1)^2-(y^2)^2\right)^{\!\!3/4}. \label{jju}
\end{split}
\end{align}
The velocity domain  is a circular cone with barycenter at the origin of coordinates (see Fig.\ \ref{min}). Its height is equal to the diameter of the base, namely $\frac{4}{\sqrt{3}}$.
\begin{equation}
D=\Big\{{\bf v}: v_3>-1/
\sqrt{3},   v_3<\sqrt{3}- 2\sqrt{ v_1^2+v_2^2}\,\Big\}.
\end{equation}

\begin{figure}[h!]
\centering
 \includegraphics[width=7cm]{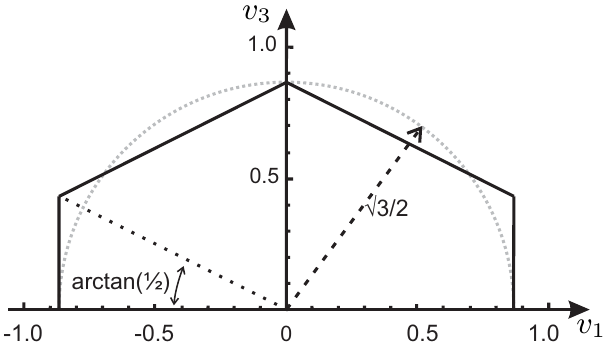}
 \caption{The two-way speed compared with the constant speed $\sqrt{3}/2$. We set $v_2=0$ since there is rotational symmetry about the third axis.} \label{twa}
\end{figure}

It can be checked that $\{ y^\alpha\}$ are indeed observer coordinates, in the sense that $\hat{y}=(1,0,0,0)$ belongs to the indicatrix and at this point $\dd \mathscr{L}=-d y^0$, $g_{\alpha \beta}=\eta_{\alpha \beta}$. The function $u$ is
\begin{equation}
u=-  \frac{2}{3^{3/8}} \left(\frac{1}{2}\!+\!\frac{\sqrt{3}}{2} \, v_3\!\right)^{\!\!\!1/4} \! \!\!\left(\!\left(\frac{\sqrt{3}}{2}\!-\!\frac{v_3}{2}\right)^2\!\!\!\!-\!\big( v_1\big)^2\!\!\!-\!\big( v_2\big)^2\!\right)^{\!\!\!3/8} \nonumber
\end{equation}
While a conic velocity domain $D$ departs very much from the sphericity of the isotropic case, it does so in a milder way with respect to the tetrahedral model. Also it must be taken into account that in most experiments only the two-way light speed is measured. This speed is the harmonic mean of the light speeds in opposite orientations, so  as Fig.\ \ref{twa} shows, the anisotropic features might appear smaller.
Let us imagine a world ruled by this type of anisotropy where the 1-2 plane could be identified at any point of the earth surface with the horizontal plane.
Although the anisotropy of the model is considerable, several experiments would not detect it, for instance if the plane x-y can be identified with the horizontal plane then it would be necessary to  tilt the plane of a Michelson-Morley apparatus in order to detect some anisotropy.


The action of the symmetry group on the coordinates $\tilde y^\alpha$ is clear. The symmetry group is a product $O^+(2,1)\times \mathbb{R}$ where the former factor is the isochronous Lorentz group while the last factor is given by the action ($\alpha \in \mathbb{R}$)
\begin{equation} \label{hof}
\tilde y^3\mapsto e^{3\alpha} \tilde y^3, \quad  (\tilde y^0, \tilde y^1,\tilde y^2)  \mapsto e^{-\alpha} (\tilde y^0, \tilde y^1,\tilde y^2).
\end{equation}
Using the change of coordinates (\ref{bot})
it is easy to write the general boost $K=S^{-1} B E S$, where $S$ is the transformation (\ref{bot}) and $B\times E$ is an element of $O^+(2,1)\times \mathbb{R}$ where $B$ is the usual boost parametrized  with a vector  $\vec{\beta}=(\beta_1,\beta_2)$ and $\gamma:=1/\sqrt{1-\beta^2}$. The matrix which sends $(y^0,y^1,y^2,y^3)^\top$ to $(y'^0,y'^1,y'^2,y'^3)^\top$ is
\[
\left(\!
\begin{array}{cccc}
\frac{1}{4}(3\gamma e^{3 \alpha}\!+\!1)  & -\frac{\sqrt{3}}{2}  \gamma \beta_1  & -\frac{\sqrt{3}}{2}  \gamma \beta_2 & \frac{\sqrt{3}}{4}(1\!-\!\gamma e^{-\alpha}) \\
 -\frac{\sqrt{3}}{2}  \beta_1 \gamma e^{3\alpha} & \frac{(\gamma-1) \beta_1^2}{\beta^2}\!+\!1 & \frac{ (\gamma-1) \beta_1\beta_2}{\beta^2} & \frac{\beta_1 \gamma e^{-\alpha}}{2 } \\
 -\frac{\sqrt{3}}{2}  \gamma \beta_2 e^{3\alpha} & \frac{ (\gamma-1) \beta_1 \beta_2}{\beta^2} & \frac{(\gamma-1) \beta_2^2}{\beta^2}\!+\!1 & \frac{\gamma \beta_2 e^{-\alpha}}{2 } \\
 \frac{\sqrt{3}}{4}(1\!-\! \gamma e^{3\alpha}) & \frac{\gamma\beta_1 }{2} & \frac{\gamma \beta_2}{2} & \frac{1}{4}(\gamma e^{-\alpha}\!+\!3) \\
\end{array}\!
\right)
\]
From the first column we read that the unprimed observer moves with velocity
\[v_1\!=\!-\frac{2 \sqrt{3} \beta_1 \gamma e^{3 \alpha}}{3 \gamma e^{3 \alpha}\!+\!1}; \,  v_2\!=\!-\frac{2 \sqrt{3} \beta_2 \gamma e^{3 \alpha}}{3 \gamma e^{3 \alpha}\!+\!1}; \,  v_3\!=\!\frac{\sqrt{3}(1\!\!-\!\gamma e^{3\alpha})}{3 \gamma e^{\!3\alpha}\!+\!1},
\]
with respect to the primed observer. We can express $(\beta_1,\beta_2,\alpha)$ in terms of $(v_1,v_2,v_3)$ as follows
\begin{align*}
\vec\beta&=-2\frac{\vec v}{\sqrt{3}-v_3},\\
{\alpha}&=\frac{1}{3}\log\left( \frac{\sqrt{(\sqrt{3}-v_3)^2-4 (v_1^2+v_2^2)}}{3 v_3+\sqrt{3}}\right).
\end{align*}
In order to obtain the velocity ${\bm \xi}$ of the primed observer with respect to the unprimed observer one can consider the first column of the inverse matrix or pass from $(v_1,v_2,v_3)$ to the group parameters $(\beta_1,\beta_2,\alpha)$, invert their signs and then calculate the corresponding value of the velocities. As a result
\[
\xi_3=\sqrt{3} \,\frac{v_1^2+v_2^2-v_3  \left( v_3-\sqrt{3}\right)}{v_1^2+v_2^2+(2v_3+\sqrt{3})(v_3-\sqrt{3})},
\]
which shows at once that ${\bm \xi}\ne -{\bf v}$, an effect due to the anisotropy of the space.
The analysis simplifies considerably for frames related with  $\vec \beta=0$. We have $e^{3 \alpha}=\frac{\sqrt{3}-v}{3v+\sqrt{3}}$, thus since $\alpha$ is an additive parameter, the law of addition of velocities along the third axis is
\begin{equation}
w=\frac{u+v+2  u v/\sqrt{3}}{1+uv}.
\end{equation}
Observe that if $u=-v$ it is not true that $w=0$. This fact means that boosting forward and then backward by the same `velocity' does not bring us back to the original frame. This is an anisotropic effect  not present in Special Relativity. In order to return to the same frame  we have to choose $u=-\frac{v}{1+2v/\sqrt{3}}$ which gives the velocity of the primed observer with respect to the unprimed observer. The law  of addition of velocities does not change if we pass from the `passive' to the `active' velocities namely whether $u,v,w$ represent the velocity of the boosted frame with respect to the original one or conversely, provided we stick to the same interpretation for all the velocities.

Also observe that if $u=\sqrt{3}$ or $u=-1/\sqrt{3}$ then the same holds for $w$ irrespective of the value of $v$. This fact is expression of the invariance of the light cone.
Finally, observe that boosts along the third axis do not affect the transversal coordinates.

Up to symmetries the non-vanishing components of the Cartan torsion are
\[
C_{311}=C_{322}=\frac{1}{\sqrt{3}}, \quad C_{333}=-\frac{2}{\sqrt{3}} .
\]
Some components of the Cartan curvature in observer coordinates can be read from the next expansion \cite{minguzzi15e} 
\begin{align*}
u({\bf v})=& \, o(\vert {\bf v}\vert^4)-1+\frac{{\bf v}^2}{2} +\frac{v^3}{\sqrt{3}} \Big[(v^1)^2+(v^2)^2-\frac{2}{3} (v^3)^2\Big]\\
&+\frac{1}{24} \left[2 \left(4 (v^3)^4+\left((v^1)^2+(v^2)^2\right)^2\right) +3({\bf v}^2)^2\right].
\end{align*}
%
%

%

Let us consider the dual formulation. Since the Finsler Lagrangian is given by (\ref{lod})
the Finsler Hamiltonian is
\begin{equation}
\mathscr{H}=-\frac{2}{3^{3/4}} (\tilde p_3)^{1/2} \big[(\tilde p_0)^2-(\tilde p_1)^2-(\tilde p_2)^2\big]^{3/4}.
\end{equation}
In observer coordinates it reads
\begin{align}
\begin{split}
\mathscr{H}&=-\frac{2}{3^{3/4}}  \Big(-\frac{1}{2}\, p_0+\frac{\sqrt{3}}{2}\, p_3\Big)^{1/2} \\&\qquad \quad \Big(\Big(\frac{\sqrt{3}}{2}\, p_0+\frac{1}{2}\, p_3\Big)^2-( p_1)^2-( p_2)^2\Big)^{3/4}.
\end{split}
\end{align}
It  does not seem possible to find a simple analytic expression for the Hamiltonian $u^*$, nevertheless we found that its Taylor expansion is  
\begin{align*}
u^*({\bf p})=& \sqrt{1+{\bf p}^2} \!-\!\frac{p_3}{\sqrt{3}} \Big[(p_1)^2\!+\!(p_2)^2\!-\frac{2}{3} (p_3)^2\Big]\!+\!o(\vert {\bf p}\vert^3),
\end{align*}
which gives the dispersion relation for  this model.

\begin{remark}
Bogoslovsky proposed an anisotropic Lagrangian intended to depart minimally from the isotropic case \cite{bogoslovsky77,bogoslovsky94}.
Its study was then revived with the proposal of the Very Special Relativity theory \cite{cohen06,gibbons07}. With a rotation of the reference frame it can be brought to the form ($b\in \mathbb{R}$ is an anisotropy parameter)
\begin{equation}
\mathscr{L}_B=-\frac{1}{2} (y^0-y^1)^{2b}[(y^0)^2-{\bf y}^2]^{1-b}.
\end{equation}
Taking the determinant of the Hessian we obtain 
\begin{align*}
\det g_{\alpha \beta}&=(b-1)^3(1+b) (y^0-y^1)^{8b} [(y^0)^2-{\bf y}^2]^{-4b}\\
&=16 (b-1)^3(1+b)  \frac{\mathscr{L}_B^4}{[(y^0)^2-{\bf y}^2]^4} ,
\end{align*}
which shows that whenever $g$ is non-degenerate it must be $\vert b\vert\ne 1$ and the determinant depends on $y$. The mean Cartan torsion does not vanish thus,
 it {\em is not} an affine sphere.  According to our previous discussion the indicatrix is not transitively preserved by a group of unimodular linear transformations, and so it does not respect the relativity principle as we defined it.
 This model for $b=1/4$ should not be confused with that given by Eq.\ (\ref{lod}).
 See \cite{weinberg95,minguzzi05e} for a discussion of the symmetries of the two factors.
\end{remark}

\section{ The general theory} \label{ftr}


In this section we consider the four-dimensional affine sphere spacetimes which satisfy the relativity principle at every point. This means that at $T_xM$ the geometry of the indicatrix belongs to one of the three types studied in the previous sections, with the difference that now $\mathscr{L}(x,y)$ might indeed  depend on $x$.

The solution of this problem
is in fact very simple and consists in introducing over each coordinate chart on $M$, a basis of one-forms $\tilde e^a=\tilde e^a_\mu(x) \dd x^\mu$, $a=0,1,2,3$, called {\em vierbeins} such that $\mu= \vert \tilde e^0\wedge \tilde  e^1\wedge \tilde  e^2\wedge \tilde e^3\vert$ is the spacetime volume form.  They provide an isomorphism between $T_xM$ and a model Lorentz-Minkowski space provided we assume that $\det \tilde e\ne 0$.
 Then the isotropic, tetrahedral anisotropic, and conical anisotropic models read respectively:
\begin{align}
\begin{split}
\mathscr{L}&=\frac{1}{2}\Big(-( \tilde e^0_\sigma(x) y^\sigma)^2+ (\tilde e^1_\sigma(x) y^\sigma)^2\\
&\qquad\qquad \qquad \qquad+(\tilde e^2_\sigma(x) y^\sigma)^2+(\tilde e^3_\sigma(x) y^\sigma)^2
\Big), \label{iso}
\end{split}\\
\mathscr{L}&=-2 \Big[\Pi_{a=0}^4(\tilde e^a_\sigma(x) y^\sigma) \Big]^{1/2},  \\
\begin{split}
\mathscr{L}&=-\frac{2}{3^{3/4}}  ( \tilde e^3_\mu(x) y^\mu)^{1/2} \Big[(\tilde e^0_\gamma(x) y^\gamma)^2-\!(  \tilde e^1_\alpha(x) y^\alpha)^2\\
&\qquad \qquad\qquad \qquad\qquad \qquad \quad-\!( \tilde e^2_\beta(x) y^\beta)^2\Big]^{3/4} . \label{con}
\end{split}
\end{align}
It is indeed clear that on each tangent space $T_xM$ we obtain the anisotropic theories studied in the previous section.

\subsection{Kinematical reformulation}
The established isomorphism between $T_xM$ and the model Lorentz-Minkowski space is largely arbitrary whenever the latter admits a symmetry group. As a consequence, it can be convenient to replace the vierbein variable with less arbitrary objects.

\begin{itemize}
\item[
\mbox{
 \includegraphics[width=0.3cm]{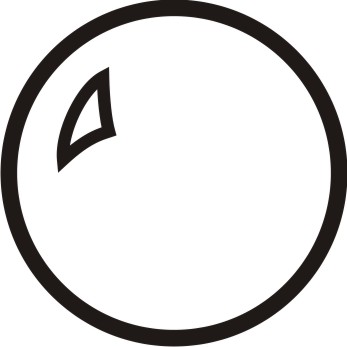}}
]
Introduced the metric $g_{\alpha \beta}(x)=\eta_{a b} \, \tilde e^a_\alpha(x) \,\tilde e^b_\beta(x)$ the isotropic model becomes
\[
\mathscr{L}=\frac{1}{2} \, g_{\alpha \beta}(x)\, y^\alpha y^\beta, \qquad \mu=\sqrt{\vert \det g_{\alpha \beta}\vert} \, \dd^4 x ,
\]
namely  the isotropic theory depends only on a Lorentzian metric.
\item[
\mbox{
 \includegraphics[width=0.3cm]{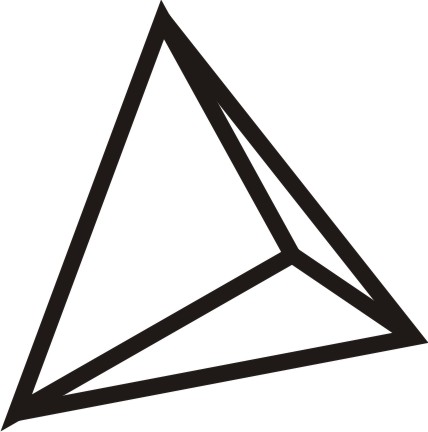}}
]
The tetrahedral anisotropic theory cannot be further simplified in the sense that one has to work with four one-forms. These forms are not completely arbitrary since $\det \tilde e\ne 0$.
\item[
\mbox{
 \includegraphics[width=0.3cm]{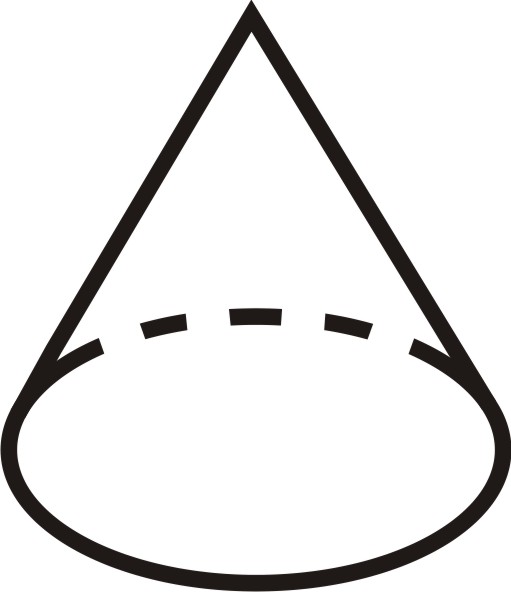}}
]  Concerning the conical anisotropic theory, let $t_\mu:=\tilde e^3_{ \mu}$, and let ${}^\xi h_{\alpha \beta}(x):=\check \eta_{a b} \, \tilde e^a_\alpha(x) \, \tilde e^b_\beta(x)$ where $\check \eta_{ab}= \eta_{ab}$ for $a,b\ne 3$ and zero otherwise. Evidently ${}^\xi h$ is a  degenerate metric of signature $(-,+,+,0)$. Its kernel is spanned by a vector $\xi$ such that $\tilde e^3(\xi)=1$, $ \tilde e^0(\xi)=\tilde e^1(\xi)=\tilde e^2(\xi)=0$, thus $t_\mu \xi^\mu=1$. Our notation ${}^\xi h$ is meant to remind us that ${}^\xi h$ is degenerate with kernel spanned by $\xi$.
Recalling the generalized Cauchy-Binet formula for the minors of a product of matrices, $M_{\alpha \beta}(AB)=\sum_\gamma M_{\alpha \gamma}(A)M_{\gamma \beta}(B)$, we obtain
\[
M_{\alpha \beta}({}^\xi h)=- M_{3 \alpha}(\tilde e) M_{3 \beta}(\tilde e) .
\]
Moreover, using the Laplace expansion for the determinant, and selecting the last row to calculate the expansion
\[
\det \tilde e= \sum_\mu (-1)^{\mu} t_\mu M_{3\mu}(\tilde e)  ,
\]
thus $(\det \tilde e)^2=-(-1)^{\alpha +\beta} M_{\alpha \beta}({}^\xi h) t_\alpha t_\beta$. This identity can be suggestively written
\[
(\det \tilde e)^2=(-\det {}^\xi h) {}^\xi h^{\alpha \beta} t_\alpha t_\beta,
\]
where it is understood that this expression is just a mnemonic aid to recover the above expression involving minors. Indeed, ${}^\xi h$ cannot be really inverted since it is degenerate. Finally, this theory is reduced to the Lagrangian and associated volume form
\begin{align*}
\mathscr{L}&=-\frac{2}{3^{3/4}} \big(t_\mu(x) y^\mu\big)^{1/2} \big( {}^\xi\! h_{\alpha \beta}(x)\ y^\alpha y^\beta\big)^{3/4} \!\!\!\!, \\  \mu&= \sqrt{\vert \det {}^\xi\! h\vert\,\vert {}^\xi \!h^{\alpha \beta}t_\alpha t_\beta\vert} \, \dd^4 x .
\end{align*}
where the square root appearing in the volume form is positive and ${}^\xi h$ has signature $(-,+,+,0)$.
\end{itemize}

The comparison of the theory with observation might require   a different choice of vierbeins \[\tilde e^a_\mu=M^a_{\, b} e^b_\mu,\] where $M$ is the matrix which in  the previous section accomplished the change of coordinates $\tilde y^a=M^a_{\, b} y^b$ (thus $M^a_{\, b}=\delta^a_b$ in the isotropic theory, while $M$ is just a rotation of 30 degrees in the 0-3 plane in the conic theory). In fact, whenever $e^i_\mu y^\mu\ll e^0_\mu y^\mu $ (e.g.\ because $e^0_0>0$, $e^i_0=0$ and $y^j\ll y^0$) we have that the Lagrangian of the tetrahedral or conic theories is approximated by the isotropic one. In other words, in that velocity limit the Finslerian kinematics reduces itself to   the general relativistic one.

\subsection{Dynamics}
This section shows how to construct a dynamical Lagrangian or the field equations for the  kinematical models. It can be skipped on first reading.

In order to define a dynamics we shall need an action. Fortunately, due to the affine sphere condition we have already a well defined volume form on $M$ so we need only to define a scalar Lagrangian. The traditional approach in Finsler gravity theory consists in trying to build, if not a Lagrangian, some field equations directly from the various curvatures associated to the Berwald, Cartan or Chern-Rund Finsler connections. This approach has been followed by Horvath \cite{horvath50},  Takano \cite{takano74}, Ishikawa \cite{ishikawa80,ishikawa81}, Ikeda \cite{ikeda79}, Asanov \cite{asanov85}, Miron \cite{miron87}, Rutz \cite{rutz93}, Li and Chang \cite{li07}, Vacaru \cite{vacaru10b}, Pfeifer and Wohlfarth \cite{pfeifer12}, to mention a few. The author has also explored this route \cite{minguzzi14c}. It has the drawback that the so obtained equations would depend on the fiber variables, a fact which complicates their interpretation as evolution equations.

Here we are going to construct dynamical equations which do not depend on the fiber variables and which, variationally speaking, do not introduce complications related to the integration over the non-compact indicatrix. We do not use the Finslerian curvatures but rather, construct a gauge theory from the fields which enter the definition of Finsler Lagrangian. The number and nature of these fields depend on the  model considered.

In fact, the most straightforward approach towards the dynamics of the theory consists in {\em gauging} the interior symmetry. This gauging is necessary since the Finsler Lagrangian is largely independent of the vierbein choice and so should be the dynamics.
As we mentioned, the interior groups of vierbein transformations which leave the Finsler Lagrangians (\ref{iso})-(\ref{con}) invariant are $O^+(3,1)$, $\mathbb{R}^3$ and $O^+(2,1)\times \mathbb{R}$,   respectively.
We assume the existence of a $G$-structure over $M$, where $G$ is the interior group. This hypothesis allows us to   assume the existence of a $\mathfrak{g}$-valued connection and hence of a $\mathfrak{g}$-valued curvature.

\begin{itemize}
\item[
\mbox{
 \includegraphics[width=0.3cm]{itemiso}}
] In the isotropic case we have a natural gauge invariant object, namely the spacetime metric $g_{\mu \nu}:= \eta_{ab} \tilde e^a_\mu \tilde e^b_\nu$. Thus a gauge invariant Lagrangian can be obtained from a scalar constructed from the metric. Of course,  general relativity tells us that the appropriate scalar is the Ricci scalar.
\item[
\mbox{
 \includegraphics[width=0.3cm]{itemtetra}}
] We have four 1-form variables $\tilde e^a_{\mu}$, $a=0,1,2,3$, and three 1-form Abelian connections $A^i_\mu$, $i=1,2,3$, due to the three Abelian gauge symmetries c.f.\ Eq.\ (\ref{nok}), $A'{}_\mu^i=A_\mu^i-\p_\mu \alpha_i$
\begin{align}
\tilde e'{}^0_\nu&=e^{-\sum_i\alpha_i}\tilde e^0_\nu,\\
\tilde e'{}^i_\nu&=e^{\alpha_i}\tilde e^i_\nu,
\end{align}
  where $e^0$ has charge $(q_1,q_2,q_3)=(-1,-1,-1)$,  $e^1$ has charge $(1,0,0)$, $e^2$ has charge $(0,1,0)$ and $e^3$ has charge $(0,0,1)$. We introduce a covariant derivative which takes into account these  charges
\begin{align*}
D_\mu \tilde e^0_\nu&=\p_\mu \tilde e^0_\nu-(\sum_iA^i_\mu)\tilde e^0_\nu,\\
D_\mu \tilde e^i_\nu&=\p_\mu \tilde e^i_\nu+A^i_\mu \tilde e^i_\nu, \quad i=1,2,3.
\end{align*}

These covariant derivatives are left invariant under the gauge transformation.
The vierbeins $e^\nu_a$ have opposite charges so that an upper interior index brings the opposite charge of a lower interior index and the interior contractions are uncharged.

Observe that we have four linearly independent 1-forms which can be arbitrarily rescaled though gauge transformations provided the volume form is left invariant. Dually, we have four linearly independent vectors which can be arbitrarily rescaled provided their wedge product is left invariant. These vierbeins determine at each point four preferred directions but no preferred scale along those directions.
It is a kind of geometry slightly more relaxed than Weitzenb\"ock's. There the connection would be obtained imposing the parallel translation of the vierbein field $\nabla^W_\alpha \tilde e^a_\mu=\p_\alpha \tilde e^a_\mu-\Gamma^W {}_{\mu \alpha}^\sigma \tilde e^a_\sigma=0$, thus $\Gamma^W {}_{\mu \alpha}^\sigma=\tilde e_a^\sigma \p_\alpha \tilde e^a_\mu $, while here we have to replace ordinary derivatives with gauge derivatives thus
\[
\empty \qquad \quad {}^{
 \includegraphics[width=0.15cm]{itemtetra}} \nabla_\alpha \tilde e^a_\mu=D_\alpha \tilde e^a_\mu\!-\!\Gamma_{\mu \alpha}^\sigma \tilde e^a_\sigma=0, \ \ \Rightarrow \ \ \Gamma_{\mu \alpha}^\sigma=\tilde e_a^\sigma D_\alpha \tilde e^a_\mu.
\]
The connection coefficients $\Gamma$ determine a linear connection $\nabla$ from which we can construct the torsion tensor
\[
T^\alpha_{\mu \nu}=\Gamma^\alpha_{\nu \mu}-\Gamma^\alpha_{\mu \nu}=\tilde e_a^\alpha D_\mu \tilde e^a_\nu-\tilde e_a^\alpha D_\nu \tilde e^a_\mu
\]
and the curvature $R^\alpha_{\ \beta \gamma \delta}$. Thus, introducing the Abelian curvatures $F^i_{\mu\nu}=\p_\mu A^i_\nu-\p_\nu A^i_\mu$, the most general action for this theory is
\[
S=\int f(R,T,{F},\tilde e)\det (  \tilde e^a_\mu) \dd x,
\]
where with $\tilde e$ we mean the vierbeins or their dual. It should be observed that contrary to the isotropic theory we do not have an interior metric $\eta_{ab}$ which through $g^{\mu \nu}=\tilde e^\mu_a \eta^{ab}\tilde e^\nu_b$ could allow us to  contract lower spacetime indices. Furthermore, $R$, $T$, $F$ are predominant in the lower indices so the construction of  a scalar appears non-trivial. Some interesting scalars are
\[
\frac{\vert \det(R_{(\alpha \beta)})\vert^{1/2}}{\det(\tilde e^a_\nu)}, \quad
 \frac{\textrm{Pf}(F_{\alpha \beta}^i)}{\det(\tilde e^a_\nu)},
\]
where $R_{(\alpha \beta)}$ is the symmetrized Ricci tensor and $\textrm{Pf}$ is the Pfaffian. The latter choice gives an action term of topological origin while the former choice is inspired by Eddington's purely affine action \cite{poplawski07}. If $B^{\alpha \beta}$ denotes the transpose of the cofactor matrix of $R_{(\alpha \beta)}$, namely the matrix such that $B^{\alpha \beta}R_{(\beta \gamma)}=\det (R_{(\alpha \beta)}) \delta^\alpha_\gamma$, then particularly interesting is the action
\[
\empty \quad\quad \ \ \int \!\left(\!\vert \det(R_{(\alpha \beta)})\vert^{1/2}\!+\!\sum_i \!c_i F^i_{\alpha \beta} F^i_{\gamma \delta} \frac{B^{\alpha \gamma} B^{\beta \delta}}{ [\det  (\tilde e^a_\alpha)
]^{3}}\right)  \dd^4 x,
\]
where $c_i$ are coupling constants.
%

Some other possibilities are offered by the tensoriality of the following uncharged object $t^{\alpha \beta \gamma \delta}:= \tilde e_0^\alpha \tilde e_1^\beta \tilde e_2^\gamma \tilde e_3^\delta$. Then other examples of scalars which might enter the construction of a Lagrangian  are
\[
 \quad R_{\alpha \beta} R_{\gamma \delta}t^{\alpha \beta \gamma \delta}, \quad F^i_{\alpha \beta} F^i_{\gamma \delta} t^{\alpha \beta \gamma \delta}, \quad R_{\alpha \beta} F^i_{\gamma \delta}t^{\alpha \beta \gamma \delta},
\]
or various combinations in the fourth power of the torsion e.g.\
\[
T^\alpha_{\beta \gamma} T^{\gamma}_{ \nu \alpha} T^{\delta}_{\eta \rho} T^\rho_{\mu \delta} t^{\beta \nu \eta \mu} .
\]
Finally, there is the possibility of writing directly field equations of non-variational origin by equating equally charged terms.
\item[
\mbox{
 \includegraphics[width=0.3cm]{itemcone}}
] In the conic theory the $O^+(2,1)$-gauge invariance can be accomplished constructing the Lagrangian from the  $O^+(2,1)$-gauge invariant fields $t_\mu$ and ${}^\xi h_{\alpha \beta}$. Additionally, we have a gauge field $A_\mu$ due to the Abelian gauge symmetry cf.\ Eq.\ (\ref{hof})
\begin{align}
A'_\mu&=A_\mu-\p_\mu \alpha,\\
t_\mu'&=e^{3\alpha}t_\mu,\\
{}^\xi \!h_{\mu \nu}'&=e^{-\alpha} \,{}^\xi\! h_{\mu \nu},
\end{align}
namely $t$ has charge $3$ while ${}^\xi h$ has charge $-1$. The vector $\xi$ has change $-3$.

The pair $(t_\mu, {}^\xi \!h_{\alpha \beta})$, where $\xi$ spans the kernel of ${}^\xi \!h_{\alpha \beta}$ and $t_\mu \xi ^\mu=1$,  can be easily shown to be equivalent to a triple $(t_\mu, h^{\alpha \beta}, \xi^\nu)$, where $h^{\alpha \beta}$ is a contravariant metric of nullity one, $h^{\alpha \beta} t_\beta=0$, and  ${}^\xi \!h^\alpha_\beta:=h^{\alpha \mu} \, {}^\xi \!h_{\mu \beta}=\delta^\alpha_\beta-\xi^\alpha t_\beta$ is the projector on $\ker t $ determined by the spitting $T_xM=(\ker t\oplus \langle \xi \rangle)\vert_x$.

The tensor $h^{\alpha \beta} $ does not bring the $\xi$ label because it is, in a well defined sense, independent of it. In fact, it really depends only on ${}^\xi \!h_{\alpha \beta}\vert_{\ker t}$. This metric is non-degenerate thus it has an inverse $({}^\xi \!h\vert_{\ker t})^{-1}$  which acts as a bilinear form on $\ker t^*$. But any element of  $\ker t^*$ can be regarded as an equivalence class of forms, any two forms being equivalent if they differ by a term proportional to $t$. As a consequence $({}^\xi \!h_{\alpha \beta}\vert_{\ker t})^{-1}$ can be represented by a contravariant metric  which annihilates $t_\beta$, this is $h^{\alpha \beta}$.

Observe that $h^{\alpha \beta}$ has charge 1.  The reader acquainted with the geometrical formulation of the Newtonian gravitational theory will recognize its main geometric ingredients \cite{kunzle72,malament12} with three relevant differences (a) the metrics $h^{\alpha \beta}$ and ${}^\xi\! h_{\alpha \beta}$ have signature $(-,+,+,0)$ rather than $(+,+,+,0)$, (b) the fields are charged, (c) the dynamics depends on a `non-relativistic matter' field $\xi$.

Let us construct a dynamics which is reminiscent of Newtonian gravity.
We introduce a derivative which takes into account the charges
\begin{align*}
D_\mu t_\nu&=\p_\mu t_\nu+3 A_\mu t_\nu,\\
D_\mu  h^{\alpha \beta}&=\p_\mu h^{\alpha \beta}+A_\mu h^{\alpha \beta},\\
D_\mu \xi^\nu&=\p_\mu \xi^\nu-3 A_\mu \xi^\nu.
\end{align*}
Next we introduce an affine connection  $\nabla$ through its coefficients $\Gamma^\alpha_{\mu \nu}$ and impose that the fields $(t_\alpha, h^{\mu \nu})$  be covariantly constant with respect to the gauged covariant derivative
\begin{align}
\empty \qquad \ \  {}^{\includegraphics[width=0.15cm]{itemcone}}\nabla_\mu t_\nu&=\nabla_\mu t_\nu\!+\!3 A_\mu t_\nu=D_{\mu} t_\nu-\!\Gamma^\alpha_{\nu\mu} t_\alpha=0, \label{pom}\\
\begin{split}
{}^{\includegraphics[width=0.15cm]{itemcone}}\nabla_\mu h^{\alpha \beta}&=\nabla_\mu h^{\alpha \beta}+\!A_\mu h^{\alpha \beta}\\
&=D_\mu h^{\alpha \beta}\!+\!\Gamma_{\sigma   \mu}^\alpha h^{\sigma \beta}\!+\!\Gamma^{\beta}_{\sigma \mu} h^{\alpha
\sigma}=0. \label{pon}
\end{split}
\end{align}
The former equation implies that the torsion $T^\alpha_{\mu \nu}:=\Gamma^\alpha_{\nu \mu}-\Gamma^\alpha_{\mu \nu}$ satisfies
\[
T^\alpha_{\mu \nu} t_\alpha
=(\dd t+3 A \wedge t)_{\mu \nu}=0,
\]
thus the connection is torsionless only if $\ker t$ is integrable. We shall assume that the connection is torsionless. Defined the curvature $F_{\mu \nu}=\p_\mu A_\nu-\p_\nu A_\mu$, the previous equations imply $F\wedge t=0$, namely the `magnetic' components vanish and so $F$ is purely `electric'.

Observe that the light cone includes a distinguished flat boundary which provides us with a distribution of hyperplanes $\ker t$ over the manifold. Since the distribution is integrable we have a natural foliation which can be interpreted as a global absolute notion of simultaneity. Over each slice we have a Lorentzian metric, thus  the spacetime $M$ is foliated by a one-parameter family of Lorentzian manifolds. Given a curve $x\colon I \to M$, $s\to x(s)$, such that $ t_\mu \frac{\dd x^\mu}{\dd s}>0$ (i.e.\ classically timelike) the integral $\int t_\mu \frac{\dd x^\mu}{\dd s}\dd s$ cannot represent the time of the particle since $t_\mu$ is not gauge invariant. This is an important difference with respect to the Newtonian theory.
 The meaningful proper time over the trajectory is that calculated via the Finsler Lagrangian: $\int \sqrt{-2\mathscr{L}(x,x')} \dd s$.
 Curiously, as we shall clarify in a moment, the conic theory  mingles a sort of formally non-relativistic field dynamics together with a  relativistic notion of proper time.

Let us raise indices with $h^{\alpha \beta}$.
As in Newton-Cartan theory \cite{kunzle72,duval78} we consider connections of the form (observe that we took into account the Abelian gauge symmetry)
\begin{align*}
\Gamma_{\alpha \beta}^\mu &=h^{\mu \sigma} \frac{1}{2}\left(D_\beta {}^\xi \!h_{\alpha \sigma}+\!D_\alpha {}^\xi \!h_{\sigma \beta}\!-\!D_\sigma {}^\xi \!h_{\alpha \beta}\right)\\
&\quad + D_{(\alpha} t_{\beta)} \xi^\mu+ t_{(\alpha} \Omega_{\beta  ) \sigma} h^{\sigma \mu} .
\end{align*}
where $\Omega_{\alpha \beta}= - 2 \,{}^\xi\!h_{\gamma [\alpha} \, {}^{\includegraphics[width=0.15cm]{itemcone}}\nabla_{\beta]} \xi^\gamma$ vanishes if and only if $\xi$ is geodesic and twist-free, $\xi^\mu\,  {}^{\includegraphics[width=0.15cm]{itemcone}}\nabla_\mu \xi^\nu=0$,  $  {}^{\includegraphics[width=0.15cm]{itemcone}}\nabla^{[\mu} \xi^{\nu]}=0$. Observe that the connection is uncharged.

Mimicking Newton-Cartan theory, the vacuum dynamics for $h^{\alpha \beta}$ and $A_\mu$ can be assigned to be
\begin{equation}
R_{\alpha \beta}=0, \quad \nabla^{\beta} F_{\alpha \beta}= 0.
\end{equation}
The vector field $\xi$ could be assigned a dynamics  formally analogous to that of a non-relativistic fluid.

Of course, completely different dynamics could have been considered, e.g. in those cases in which $\Gamma$ has torsion.
 In fact, many scalars can be built from the torsion and curvature of $\Gamma$. In order to contract lower indices one could use the tensoriality of the object \[(\det {}^\xi\! h) {}^\xi\! h^{\mu \nu}/[\vert \det {}^\xi\! h\vert\,\vert {}^\xi\! h^{\alpha \beta}t_\alpha t_\beta\vert].\]
\end{itemize}

These considerations were aimed at illustrating the possibility of defining a dynamics for the Finslerian kinematical theories previously introduced. In the next section we shall show that it is not necessary to impose some dynamical equations and to solve them in  order to select physically interesting affine sphere spacetimes. Indeed, these spaces will be uniquely selected from the imposition of an appropriate general relativistic limit. These notable spacetimes might then help to select the correct field equations.

%

%

\subsection{Notable 
affine sphere spacetimes}

We can construct some first examples of general relativistic affine sphere spacetimes which satisfy the relativity principle. We shall impose that at every point the spacetime is conic anisotropic obtaining conic anisotropic generalizations of the Kerr-Schild, Schwarzschild, Kerr, Taub, FLRW metrics. A test particle slowly moving on these spacetimes with respect to their natural stationary observer would behave as in the corresponding Lorentzian spacetimes of general relativity. I have not been able to obtain similarly good results for the tetrahedral theory.

\subsection{Conic anisotropic generalization of the Kerr-Schild metric}
 We recall that the fiber coordinate is defined by $y^\mu=\dd x^\mu\colon T_xM\to \mathbb{R}^{n+1}$. In this section we might revert to the notation $\dd x^\mu$ for the fiber coordinate.
Let  $f\colon U\to \mathbb{R}$, $\mu\colon U \to (0,2\pi)\backslash\{\pi/2,\pi,3\pi/2\}$, $U\subset M$, be  functions and let $k=k_\alpha \dd x^\alpha=\dd t+ k_x \dd x+k_y \dd y+k_z \dd z$  be a 1-form field on the  same coordinate patch $U$. Let us define
\begin{align*}
\beta&=\sqrt{(1-f) + f k_z^2},\\
\omega_\perp&=k_x \dd x+k_y\dd y,\\
\omega_{t}&=\dd  t- \frac{f}{1-  f} \, (k_x \dd x+k_y \dd y+k_z \dd z),\\
\omega_z&= \frac{1}{\beta} \left(\dd z+ \frac{f}{1-  f }\, k_z (k_x \dd x+k_y \dd y+k_z \dd z)\right) .
\end{align*}
Let us consider the Finsler Lagrangian
\begin{align} \label{gen}
\begin{split}
\mathscr{L}&=-\frac{1-f}{2 (\cos^2 \!\mu)^{\cos^2 \!\mu} (\sin^2\!\mu)^{\sin^2 \!\mu}} \,
\left(\!\left(\sin \mu \,\omega_{ t}\!+\!\cos \mu \,\omega_z\right)^2\right)^{\! \sin^2\!\mu}
\\
& \!\!\!\!
\left(\!\left(\cos \mu \,\omega_{ t}\!-\!\sin \mu\, \omega_z\right)^2\!-\!\frac{1}{1\!-\!f} \left(\dd x^2\!+\!\dd y^2\!+\!\frac{f}{\beta^2}\,\omega_\perp^2\right)\!\right)^{\!\! \cos^2 \!\mu}
\end{split}
\end{align}
This expression is left invariant if we change the orientation of $z$,  $x$ with $y$, and the sign of $\sin \mu$, thus $\mu$ can be assumed in the range $(0,\pi)$ with no loss of generality.

Its limit for large distances (large $\max(\vert x_i\vert)$) is
\begin{align}
\mathscr{L}_\infty&=-\frac{1}{2 (\cos^2 \! \bar \mu)^{\cos^2 \!\bar \mu} (\sin^2\! \bar\mu)^{\sin^2 \!\bar \mu}}
\left( \!\left(\sin \bar \mu \,\dd  t\!+\!\cos \bar \mu \,\dd z\right)^2\!\right)^{\!\sin^2\! \bar\mu}
 \nonumber \\ &\qquad
\left(\left(\cos \bar \mu \,\dd  t\!-\!\sin \bar \mu \, \dd z\right)^2\!- \!\dd x^2\!-\!\dd y^2\right)^{\cos^2 \! \bar\mu} ,
\end{align}
provided for every $\alpha,
 \beta$, we have $f k_\alpha k_\beta\to 0$ and $\mu \to \bar\mu$ in that limit.

If $\mu=\bar \mu$ is a constant throughout $M$ then $\mathscr{L}$ is modeled on the same Lorentz-Minkowski space $\mathscr{L}_\infty$ at every point.

At every point $x\in M$ the vector $\hat y=(\frac{1}{\sqrt{1-f}},0,0,0)$ belongs to the indicatrix and so provides an observer vector field which will be of particular interest whenever $(M,\mathscr{L})$ is stationary, that is, independent of time.

For low velocities with respect to $\hat y$, $y^i\ll y^0$, and for every function $\mu(x)$, the Lagrangian reduces itself to the Kerr-Schild metric
\[
\dd s^2=-\dd  t^2+\dd x^2+\dd y^2+\dd z^2+f k_\alpha k_\beta \dd x^\alpha \dd x^\beta.
\]
Under the assumption $f k_\alpha k_\beta\to 0$  it is asymptotic to  the Minkowski metric which is indeed the  low velocity limit of $\mathscr{L}_\infty$.

For $\mu=\pi/6$ namely with
\begin{align}
\mathscr{L}&=-\frac{2 (1-f) }{3^{3/4}}\,
\left(\!\left(\frac{1}{2} \,\omega_{ t}+\frac{\sqrt{3}}{2} \,\omega_z\right)^2\right)^{\! 1/4}
 \\ &\quad
\left(\!\left(\frac{\sqrt{3}}{2}\,\omega_{ t}\!-\!\frac{1}{2}\, \omega_z\right)^2\!\!\!-\!\frac{1}{1\!-\!f} \left(\dd x^2+\dd y^2\!+\!\frac{f}{\beta^2}\,\omega_\perp^2\right)\!\right)^{\!\!\!3/4} \nonumber
\end{align}
the indicatrix is a Calabi product of affine spheres, thus it is itself an affine sphere and hence its mean Cartan torsion vanishes. Its asymptotic limit and model Lorentz-Minkowski space is
\begin{align}
\begin{split}
\mathscr{L}_\infty&=-\frac{2  }{3^{3/4}}
\left( \left(\frac{1}{2} \,\dd  t+\frac{\sqrt{3}}{2} \,\dd z\right)^{\!2}\right)^{\!\!1/4}  \\
 &\qquad \left(\left(\frac{\sqrt{3}}{2}\,\dd  t-\frac{1}{2}\, \dd z\right)^{\! 2}\!- \dd x^2-\dd y^2\!\right)^{\!\!3/4}
\end{split}
\end{align}

If $\mu$ is different from this special value the mean Cartan torsion does not vanish. Indeed, a calculation at the observer $\hat y$ gives
\begin{equation}
I_\alpha(\hat y)=\frac{2  (3-4 \cos^2 \!\mu)}{\beta\sqrt{1\!-\!f}\,\cos \mu\sin \mu}\,\big(0,f k_1k_3,f k_2k_3, \beta^2\big) .
\end{equation}
 Now, for any chosen $\mu(x)$ we can obtain from (\ref{gen}) the Finslerian conic anisotropic version of many general relativistic metrics. For instance, for the Kerr-Newman metric in Kerr-Schild Cartesian coordinates \cite{hawking73} we set for some constants $m>0,a,q$
\begin{align*}
k_\alpha&=\big(1,\frac{rx+ay}{a^2+r^2}, \frac{ry-ax}{a^2+r^2},\frac{z}{r}\big),\\
f&=\frac{2 m r^3-q^2 r^2}{a^2 z^2+r^4},
\end{align*}
where $r(x,y,z)$ is determined implicitly, up to a sign, by the requirement that $k$ be null, namely
\[
\frac{x^2+y^2}{a^2+r^2}+\frac{z^2}{r^2}=1.
\]
Similarly, the Kerr-de Sitter metric can be obtained from $k$ and $r$ as above with $a=0$, by setting
\[
f=\frac{2m}{r}+\frac{\Lambda}{3} r^2.
\]
For the Schwarzschild metric ($a=\Lambda=0$) it can be convenient to introduce cylindrical coordinates $(z,\rho,\varphi)$, pass to the Schwarzschild time $t_S$ through \[t = t_S+ 2 m \ln \vert \frac{r}{2m}-1\vert ,\] in such a way that $\omega_t=\dd t_S$,  set $r=\sqrt{z^2+\rho^2}$ and set for definiteness $\mu=\pi/6$, then
\begin{widetext}
\begin{equation}
\begin{split}
\mathscr{L}&=-\frac{2 (1-\frac{2m}{r}) }{3^{3/4}}\,
\left(\left(\frac{1}{2} \,\dd t_S+\frac{\sqrt{3}}{2} \left(1-\frac{2m}{r}\right)^{\!-1} \frac{(1-2m \rho^2/r^3)\dd z+2mz\rho\, \dd \rho}{\sqrt{1-2m\rho^2/r^3}}\right)^{\!2}\right)^{\!1/4}
 \\ &\quad
\left(\!\left(\frac{\sqrt{3}}{2}\,\dd t_S\!-\!\frac{1}{2}\, \left(1-\frac{2m}{r}\right)^{\!-1}\! \frac{(1\!-\!2m \rho^2/r^3)\dd z+2mz\rho\, \dd \rho}{\sqrt{1-2m\rho^2/r^3}}\right)^{\!\!2}-\left(1-\frac{2m}{r}\right)^{\!-1} \! \left(\frac{\dd \rho^2}{1-2m\rho^2/r^3}+\rho^2 \dd \varphi^2\right)\right)^{\!\!3/4}.
\end{split}
\end{equation}
\end{widetext}
%
%

The metric can be written using Boyer-Linquist coordinates $(r,\theta,\varphi)$ defined by
\begin{align*}
x+iy&=(r+ia) \sin \theta \exp \, i\left(\varphi+a \int\frac{\dd r}{r^2-2m r +a^2}\right) , \\ z&=r \cos \theta, \qquad
\bar t = t+ 2 m \int\frac{ r \dd r}{r^2-2m r +a^2},
\end{align*}
by noticing  that
\begin{align*}
\omega_\perp&=\frac{(r^2-2 m r)\sin^2 \theta}{r^2-2m r +a^2} \dd r+r\sin \theta \cos \theta \dd \theta \\&\quad-a \sin^2 \theta \dd \varphi,\\
\omega+\frac{z\dd z}{r}&=\left(1-\frac{a^2\sin^2 \theta}{r^2-2m r +a^2} \right)\dd r-a\sin^2 \theta \dd \varphi,\\
\dd x^2+\dd y^2&=\frac{r^2 \sin ^2(\theta ) \left(a^2+(r-2 m)^2\right)}{\left(r^2-2m r+a^2 \right)^2} \dd r^2 \\
&\quad+(r^2+a^2) \sin^2\theta \dd \varphi^2+(r^2+a^2) \cos ^2 \theta \dd \theta^2 \\
&\quad+\frac{4 a m r \sin ^2\theta }{r^2-2m r+a^2 } \dd r \dd \varphi+2 r \cos \theta \sin \theta \dd r \dd \theta .
\end{align*}
The final expression is not particularly illuminating, however, it shows that the Finsler Lagrangian has  Killing vectors $\p_t$, $\p_\phi$. We have
\begin{align*}
\alpha&=1-\frac{2 m r}{ r^2+a^2 \cos^2 \theta},\\
\beta&=\sqrt{1 -\frac{2 m  \sin^2 \theta}{r}},\\
\omega_\perp&=\frac{(r^2-2 m r)\sin^2 \theta}{r^2-2m r +a^2} \dd r+r\sin \theta \cos \theta \dd \theta \\&\quad -a \sin^2 \theta \dd \varphi , \\
\omega_{\bar t}&=\dd  t +  \frac{2m r  \sin^2 \theta}{r^2-2 m r+a^2 \cos^2 \theta}  \, a\dd \varphi ,\\
\omega_z&=\frac{1}{\beta}\Big\{ \frac{\left(a^2+r^2\right) \cos \theta }{r^2-2 m r +a^2} \dd r  - r\sin \theta \dd \theta\\& \ \ \qquad- \frac{2m r  \cos \theta\sin^2 \theta}{r^2-2mr +a^2 \cos^2 \theta}\, a \dd \varphi  \Big\} ,\\
\dd x^2+\dd y^2&=\sin^2 \theta \dd r^2+r^2 \sin^2\theta \dd \phi^2+r^2 \cos ^2 \theta \dd \theta^2\\&\quad+2 r \cos \theta \sin \theta \dd r \dd \theta.
\end{align*}
 The Finsler Lagrangian becomes
\begin{align}
\mathscr{L}&=-\frac{2 \alpha }{3^{3/4}} \left(\frac{1}{2} \,\omega_{\bar t}+\frac{\sqrt{3}}{2} \,\omega_z\right)^{\!1/2} \\& \quad \left(\left(\frac{\sqrt{3}}{2}\,\omega_{\bar t}-\frac{1}{2}\, \omega_z\right)^{\!2}\!\!\!-\!\frac{1}{\alpha} \left(\dd x^2+\dd y^2+\frac{2 m }{ r\beta^2}\,\omega_\perp^2\right)\right)^{\!\!\!3/4} \nonumber
\end{align}
The low velocity limit gives the Kerr metric in Boyer-Linquist coordinates. For $a=0$, $\omega_{\bar t}=\dd t$,  the low velocity metric is Schwarzschild's and $t$ is the Schwarzschild's time.

\subsection{A cosmological model}
In this section we shall construct the conic anisotropic versions of the FLRW metrics with $k=1$ or $k=0$. We shall also obtain the conic anisotropic version of the Taub solution.  For $k=1$ the idea is to regard the $S^3$ space section as a Hopf fibration and to orient the anisotropic direction of the conic anisotropy along the Clifford parallels, that is, along the fibers.

\subsubsection{The Hopf bundle}

Let us first recall the construction of the Hopf fibration. This introduction will also serve to fix the notation. Let an element of $SU(2)$ be parametrized as follows
\[
w=\begin{pmatrix}
 z_0 &- \bar{z}_1\\
 z_1& \bar{z}_0
\end{pmatrix}, \qquad \vert z_0\vert^2+\vert z_1\vert^2=1.
\]
This expression clarifies that $SU(2)$ is diffeomorphic to $S^3$.
Let us denote with $\sigma_i$ the Pauli matrices
\[
\sigma_1=
\begin{pmatrix}
0 & 1\\1&0
\end{pmatrix}, \quad \sigma_2=\begin{pmatrix}
0 & -i\\i&0
\end{pmatrix}, \quad \sigma_3=\begin{pmatrix}
1 & 0\\0&-1
\end{pmatrix},
\]
and let $\tau_k=i\sigma_k/2$ be the generators of the Lie algebra $\mathfrak{su}(2)$,
\[
[\tau_i,\tau_j]= \varepsilon_{ijk} \tau_k.
\]
Every element of $SU(2)$ is also a linear combination of the identity and $\tau_k$. It will be useful to recall the identity
\[
\sigma_i \sigma_j = i \varepsilon_{ijk} \sigma_k + \delta_{ij} \, I,
\]
and that $\det \sigma_i=-1$.
Let us define the map over $SU(2)$
\[
\pi(w)=2w \tau_3 w^{\dagger},
\]
some algebra shows that
\[
\pi(w)=2w \tau_3 w^{\dagger}=i\begin{pmatrix}  a& \bar b \\  b & -a\end{pmatrix}=
\begin{pmatrix}  i a& - \overline{i b} \,\\ i b & \overline{ i a}\end{pmatrix},  \]
where  $a=\vert z_0\vert^2-\vert z_1\vert^2 \in \mathbb{R}$, and  $b= 2  z_1 \bar z_0 \in \mathbb{C}$.
Observe that $\pi(w)$ belongs to $SU(2)\cap \mathfrak{su}(2)$ thus $\det \pi(w)=1$ which reads $a^2+\vert b\vert^2=1$. We conclude that $\pi(w)\in S^2$.

The group $SU(2)$ admits a subgroup isomorphic to $U(1)$ given by the matrices of the form
\[
\rho(\varphi)=\begin{pmatrix} e^{i\varphi}&0\\0&e^{-i\varphi} \end{pmatrix},
\]
which is generated by $\tau_3$. Its right action on $SU(2)$ can be defined through
\begin{align*}
U(1)\times SU(2)&\to SU(2),\\
(w,\rho(\varphi))&\mapsto w \rho(\varphi).
\end{align*}
Since $\rho(\varphi)$ commutes with $\tau_3$
\[
\pi(w \rho(\varphi))=2w\rho(\varphi) \tau_3 \rho(\varphi)^{-1} w^{-1}=2w \tau_3 w^{-1}=\pi(w).
\]
Thus the projection $\pi\colon S^3\to S^2$ has fiber $S^1$. This is the Hopf fiber bundle. Let $u\in S^2$, namely let $u$ be a matrix of the form $2w\tau_3 w^\dagger$ for $w\in SU(2)$, if $h \in SU(2)$, $h u h^{-1}=\pi(h w)\in S^2$ thus $SU(2)$ acts on $S^2$ as a transformation induced from a linear transformation of $\mathbb{R}^3$. We shall see later that this is an isometry, so that $SU(2)$ acts as a rotation. This is the double covering of $SU(2)$ over $SO(3)$.

\subsubsection{Metrics over the Hopf bundle}
The idea is to construct the cone of the Finsler Lagrangian as the product between a one dimensional cone and a three dimensional irreducible cone, or equivalently the indicatrix should be the Calabi product between a zero dimensional affine sphere and an irreducible two dimensional affine sphere. We are going to construct the three dimensional cone from a Lorentzian metric on the Hopf fiber bundle. We wish to avoid coordinates as far as possible so as to make the presentation clearer. Coordinates will be introduced in the end. The (left-invariant) Maurer-Cartan form of $SU(2)$ is
\[
\theta=w^\dagger \dd w=\begin{pmatrix}
\bar z_0 \dd z_0+\bar z_1\dd z_1&-\bar z_0\dd \bar z_1+\bar z_1\dd \bar z_0\\
-z_1\dd z_0+z_0\dd z_1&z_1\dd \bar z_1+z_0 \dd \bar z_0
\end{pmatrix}
\]
It can be observed that since $\vert z_0\vert^2+\vert z_1\vert^2=1$ we have $\textrm{tr} \,\theta=0$. It can be interesting to observe that for an arbitrary $2 \times 2 $ matrix $M$ (this formula admits generalization to higher dimensions)
\[
\det M=\frac{1}{2}\det \begin{pmatrix} \textrm{tr} M & 1 \\ \textrm{tr} M^2 &  \textrm{tr} M \end{pmatrix}=
 \frac{1}{2}\left((\textrm{tr} M)^2-\textrm{tr} M^2\right),
\]
thus
\begin{align} \label{kis}
-\frac{1}{2}\textrm{tr} (\theta^2)&=\frac{1}{2}\textrm{tr}(\dd w^\dagger\dd w)=\det(\theta)\nonumber\\
&=\dd \bar z_0\dd z_0+\dd \bar z_1\dd z_1=g_{S^3}.
\end{align}
This is precisely the metric induced on $S^3$ by the Euclidean metric in $\mathbb{R}^4$ (decompose $z_0$ and $z_1$ in real and imaginary components).

Similarly, the metric induced on $S^2$ by the Euclidean metric of $\mathbb{R}^3$ is
\begin{align}\label{dis}
\begin{split}
-\frac{1}{2}\textrm{tr} \left( (\pi(w)^\dagger \dd \pi(w))^2\right)&=\dd \overline{(ia)} \dd (ia)+\dd \overline{(ib)}\dd (ib)\\ &=\dd a^2+\dd \bar b\dd b=g_{S^2}.
\end{split}
\end{align}

Since $\theta$ is $\mathfrak{su}(2)$-valued  we decompose it as follows $\theta=\tau_k \omega_k$ where $\omega_k$ are real 1-forms over $SU(2)$. Using  $\textrm{tr}(\sigma_i \sigma_j)=2\delta_{ij}$ or $\textrm{tr}(\tau_i \tau_j)=-\frac{1}{2}\delta_{ij}$  we get
\[
\omega_k=-2\textrm{tr}(\theta \tau_k).
\]
This expression shows at once that $\omega_3$ is invariant under the right action of $U(1)$, indeed let us calculate $R_a^*\omega_3$ with $a\in SU(2)$,  (observe that $R_a^*\theta=(wa)^\dagger \dd (wa)=a^\dagger \theta a$)
\begin{align*}
R^*_{a}\omega_3(X)&=-2\textrm{tr}(\theta(R_{a*} X)) \tau_3)=-2\textrm{tr}( (R^*_{a}\theta)( X)) \tau_3)\\&=-2\textrm{tr}(a^{-1}\theta(X)a \tau_3),
\end{align*}
so since $\rho(\varphi)$ commutes with $\tau_3$, $R^*_{\rho(\varphi)} \omega_3=\omega_3$. The 1-form $\omega_3$ is actually a connection for the Hopf bundle. Indeed, the vertical fundamental field is $\tau_3^*$, and by definition of $\theta$, $\theta(\tau^*_3)=\tau_3$, thus $\omega_3(\tau^*_3)=-2\textrm{tr}(\tau_3 \tau_3)=1$ (see \cite{kobayashi63} for the conditions defining a connection on a principal bundle).

There is also a $U(1)$-invariant metric, indeed,
\[
\omega_1^2+\omega_2^2=(\omega_1^2+\omega_2^2+\omega_3^2)-\omega_3^2=-2 \textrm{tr}( \theta^2)-\left(2\textrm{tr}(\theta \tau_3)\right)^2.
\]
The validity of this equation can be checked inserting $\theta=\omega_k\tau_k$ and using again $\textrm{tr}(\tau_i \tau_j)=-\frac{1}{2}\delta_{ij}$. Arguing as above $R^*_{\rho(\varphi)} (\omega_1^2+\omega_2^2)=\omega_1^2+\omega_2^2$.

As the next trace vanishes
\begin{align*}
\textrm{tr} \!\left( \pi(w)^\dagger \!\dd \pi(w)\right)
&=4\textrm{tr} \! \left[ w \tau_3 w^\dagger\!  \big( w\tau_3 (-w^\dagger\! \dd w w^\dagger) \!+\! \dd w \tau_3  w^\dagger\big)\right]\\
&=\textrm{tr}\! (\dd w w^\dagger)-\textrm{tr} ( w^\dagger\dd w)=0.
\end{align*}
we can write
\begin{align*}
-\frac{1}{2}&\textrm{tr} \!\left( (\pi(w)^\dagger \!\dd \pi(w))^2\right)=\det(\pi(w)^\dagger \dd \pi(w))=\det \!\big( \dd \pi(w)\big)\\
&=4\det (\dd w\tau_3 w^\dagger-w \tau_3 w^\dagger \dd w w^\dagger)\\
&=4\det (w^\dagger\dd w\tau_3 - \tau_3 w^\dagger \dd w )=4\det ([\theta ,\tau_3])\\
&= 4\det(-\omega_1 \tau_2+\omega_2 \tau_1)=\omega_1^2+\omega_2^2.
\end{align*}
 This result jointly with Eq.\ (\ref{dis}) shows that $\omega_1^2+\omega_2^2$ is the ($\pi$-pullback of the) canonical metric of $S^2$. Observe that the action of $SU(2)$ on $S^2$, $\pi(w)\mapsto h \pi(w) h^{-1}$ is an isometry for this metric which proves the earlier statement that $SU(2)$ is a double covering of $SO(3)$ ($h$ and $-h$ give the same map).

\begin{remark}
If one insists on using coordinates it is convenient to parametrize $SU(2)$
as follows
\[
w(\phi,\theta,\psi)=\begin{pmatrix} e^{\frac{i}{2}(\psi-\phi)} \cos (\theta/2) &  -e^{-\frac{i}{2}(\psi+\phi)} \sin (\theta/2)\\
e^{\frac{i}{2}(\psi+\phi)} \sin (\theta/2) & e^{-\frac{i}{2}(\psi-\phi)} \cos (\theta/2)
 \end{pmatrix},
\]
that is
\[
z_0=e^{\frac{i}{2}(\psi-\phi)} \cos (\theta/2), \qquad z_1= e^{\frac{i}{2}(\psi+\phi)} \sin (\theta/2),
\]
with $\phi\in [0,2\pi)$, $\psi\in [0,4\pi)$, $\theta\in [0,\pi]$ (the  angle $\psi$ can be given the domain $[0,2\pi)$ if one is interested in generating the $SO(3)$ group through the action $x'^i\sigma_i=w x^i\sigma_i w^\dagger$, however in order to generate $SU(2)$ one needs to double the domain of $\psi$ in order to generate the negated matrices.

This parametrization is particularly useful because
\[
\pi(w(\phi,\theta,\psi))=iw \sigma_3 w^\dagger=i n^k \sigma_k
\]
with $n^1=\sin \theta \cos \phi$, $n^2=\sin \theta \sin \phi$, $n^3=\cos \theta$. 
The invariants under $U(1)$-right translations are
\begin{align*}
\omega_3&=\dd \psi-\cos \theta \dd \phi,\\
\omega_1^2+\omega_2^2&= \dd \theta^2+\sin^2 \theta \dd \phi^2.
\end{align*}
The other 1-forms are
\[
\omega_1\!=\!\sin \psi \dd \theta+\cos \psi \sin \theta \dd \phi; \ \omega_2\!=\!-\cos \psi \dd \theta+\sin \theta \sin \psi \dd \phi.
\]

\end{remark}

Any metric over $S^3$ of the form $h_{ij}\omega_i\omega_j$, where $h_{ij}$ are constant coefficients,  is necessarily invariant under the left $SU(2)$ action as the forms $\omega_i$ are.
There are Riemannian metrics over $S^3$ which share additional symmetries. For instance from Eq.\ (\ref{kis})  the metric
\begin{align*}
\omega_1^2+\omega_2^2+\omega_3^2&=-2\textrm{tr} (\theta^2)=4 g_{S^3}\\
&=(\dd \psi-\cos \theta \dd \phi)^2+\dd \theta^2+\sin^2 \theta \dd \phi^2
\end{align*}
is  invariant  under the right $SU(2)$ action. This means that the isotropy group at a point, namely the subgroup which leaves a point fixed, is three dimensional a fact which implies that this space is isotropic.

In order to construct the mentioned product of cones we need a Lorentzian metric over $S^3$.
We are  interested in Lorentzian metrics over $SU(2)$ of the form
\begin{equation} \label{gi}
g=-\tilde \alpha_3^2 \omega_3^2+\alpha_\perp^2  (\omega_1^2+\omega_2^2)
\end{equation}
The forms $\omega_3$ and $\omega_1^2+\omega_2^2$ entering this metric are invariant under the $SU(2)$ left action and the  $U(1)$ right action. The metric $g$  shares similar symmetries depending on the  functions $\tilde \alpha_3$ and $\alpha_\perp$. For instance, it  respects the full symmetry if they are constant while it  respects the $U(1)$ symmetry for  $\tilde \alpha_3, \alpha_\perp\colon S^2\to \mathbb{R}$.  We are interested in the former case for it admits an additional $\tau_3$ right-rotation which tells us that the isometry subgroup which leaves a point fixed is non-trivial (not just the identity) and so that there is isotropy at least under rotations with respect to some direction. This is the direction towards which we orient the cone domain of the conic anisotropy.


%


A pointwise Calabi product and the requirement of  preservation of symmetry lead us to the next affine sphere spacetime
\begin{align}
\begin{split}
\mathscr{L}&=-\frac{2}{3^{3/4}}  \left(\!\left(\frac{1}{2} \alpha_0 \dd t+\frac{\sqrt{3}}{2} \alpha_3 \omega_3\right)^{2}\right)^{\!1/4}\\&\qquad \left(\!\left(\frac{\sqrt{3}}{2} \alpha_0 \dd t\!-\!\frac{1}{2} \alpha_3 \omega_3\right)^{\!\!2}\!-\!\alpha_\perp^2 (\omega_1^2+\omega_2^2)\right)^{\!3/4} \label{frc}
\end{split}
\end{align}
where $\alpha_3,\alpha_0, \alpha_\perp$ depend on $t$. Observe that the $U(1)$ right translations and the $SU(2)$ left translations acting on the space sections $S^3$ are symmetries for this Finsler Lagrangian. It can share additional symmetries for particular choices of  $\alpha_3,\alpha_0, \alpha_\perp$. For instance, if they are constant there is an additional $\mathbb{R}$ factor due to the time translations.

For low velocities it becomes
\[
\dd s^2=- \alpha_0^2\dd t^2+\alpha_3^2 \omega_3^2+\alpha_\perp^2 \left(\omega_1^2+\omega_2^2\right) ,
\]
which for constants $m,\ell >0$, once we set
\begin{align*}
\alpha_0^2&=U^{-1}, \qquad U(t):= \frac{\ell^2-2mt+t^2}{t^2+\ell^2} ,\\
\alpha_\perp^2&=t^2+\ell^2,\\
\alpha_3^2&= 4 \ell^2 U,
\end{align*}
gives the Taub vacuum. For $\alpha_3=\alpha_\perp =a(t)/2$, $\alpha_0=1$, it gives the FLRW metric with $k=1$
\begin{equation} \label{fdo}
\dd s^2=-\dd t^2+a^2(t) g_{S^3} .
\end{equation}
Clearly, the FLRW metric with $k=0$ can be obtained as the low velocity limit of the Finsler Lagrangian
\begin{align}
\mathscr{L}&=-\frac{2}{3^{3/4}}\left(\left(\frac{1}{2}  \dd t+\frac{\sqrt{3}}{2} a(t) \dd z\right)^{2}\right)^{1/4} \\ &\qquad\left(\!\left(\frac{\sqrt{3}}{2}  \dd t-\frac{1}{2} a(t) \dd z\right)^{\!\!2}- a^2(t) (\dd x^2+\dd y^2)\right)^{\!3/4} \nonumber
\end{align}
however, there seems to be no natural conic Finslerian generalization of the FLRW metric with $k=-1$.

\begin{remark}
It can be observed that while the FLRW Lagrangian for $k=1$, Eq.\ (\ref{frc}), has invariance group $U(1)\times SU(2)$, its low velocity limit,  Eq.\ (\ref{fdo}), has more symmetries, as it has six Killings.
This fact has to be expected on the following ground. In general, the Finsler Lagrangian captures also the kinematics of light which could be highly anisotropic, still in the low velocity limit one has that the indicatrix is approximated by a hyperboloid, which is isotropic. As a consequence, one does not see the anisotropy of velocity space but only that of spacetime and so gets more symmetries (unless the Finslerian spacetime is obtained aligning the velocity space anisotropy with that already present in its general relativistic limit as in  the Kerr example). The same phenomenon can be seen with Eq.\ (\ref{jju}) which has an eight dimensional group of symmetries while the limit for low velocities is Minkowski spacetime which has ten Killings.  \\
\end{remark}

\section{Conclusions}

In this work we have  recognized that the relativity principle is expressed by the homogeneity of the observer space (indicatrix), meaning by this its transitivity under the action of a unimodular linear group acting on the tangent space.  We have also pointed out that in four  spacetime dimensions there are only three theories which respect an exact form of the relativity principle, the velocity domain of massive particles as seen from a local observer being given by a ball, a tetrahedron or a cone, respectively. We have studied their kinematics, particularly that of the conic theory since it was not previously recognized. For each of these theories we have provided observer coordinates, namely special coordinates for which the metric becomes Minkowskian in the appropriate velocity limit.

In Sec.\ \ref{ftr} we have discussed the dynamics showing how to build consistent field equations by gauging the interior symmetries. We did not focus on particular dynamical laws. Instead, we observed that notable Finslerian spacetimes could be selected by two requirements (a) the spacetime is relativistic invariant (the indicatrix is homogeneous), (b) the low velocity limit with respect to a natural  (conformal) stationary observer returns some notable general relativistic metric. Using this approach we have been able to obtain the conic anisotropic version of the Kerr-Schild metric and through it the conic anisotropic versions of the Schwarzschild, Kerr-de Sitter and Kerr-Newman spacetimes. The generalization of the FLRW metric required a preliminary study of the Hopf bundle, but in the end we obtained the conic anisotropic versions for $k=0,1$, and as a bonus we obtained also the conic anisotropic version of Taub's spacetime.

Our study shows that other and different general relativistic theories are possible. In fact some theories might present curious hybrid features, namely the gravitational fields might admit a sort of formally non-relativistic  description while test particles might exhibit typical relativistic features, such as time dilation.


 The found geometries could possibly describe  peculiar  gravitational regions of the Universe.
 For our spacetime neighborhood a perturbative approach seems more appropriate since the local light cones are expected to depart  slightly from isotropy. Approaches which try to retain an almost general relativistic dynamics while modifying the indicatrix in a neighborhood of a (stationary) observer should  pass through a study of modified dispersion relations at the lowest order of approximation \cite{girelli07,kostelecky11,russell15}. A perturbative study respecting the geometry of affine spheres will be presented in future work.




\end{document}